\documentclass[a4paper,fleqn,usenatbib]{mnras}

\usepackage{newtxtext,newtxmath}

\usepackage[T1]{fontenc}
\usepackage{ae,aecompl}

\usepackage{graphicx}	
\usepackage{amsmath}	
\usepackage{amssymb}	
\usepackage{lscape}

\newcommand{\kep}{\textit{Kepler} }

\newcommand{\todcor}{\textsc{todcor} }

\newcommand{\jkt}{\textsc{jktebop} }
\newcommand{\jktabs}{\textsc{jktabsdim} }
\newcommand{\vfit}{\textsc{v2fit} }

\newcommand{\ms}{m~s$^{-1}$}
\newcommand{\isp}{\textsc{ispec} }

\title[HIDES spectroscopy of \kep DEBs III]{HIDES spectroscopy of bright detached eclipsing binaries
from the \kep field -- III. Spectral analysis, updated parameters, and new systems.}

\author[K. G. He\l miniak et al.]{K. G. He\l miniak,$^{1}$\thanks{E-mail: xysiek@ncac.torun.pl (KGH)}
M. Konacki,$^{1}$
H. Maehara,$^{2}$
E. Kambe,$^{3,2}$
N. Ukita,$^{2,4}$
\and
M. Ratajczak,$^{5,6}$ 
A. Pigulski,$^{6}$ and
S. K. Koz\l owski$^{1}$
\\
$^{1}$Nicolaus Copernicus Astronomical Center, Polish Academy of Sciences, ul. Rabia\'{n}ska 8, 87-100 Toru\'{n}, Poland\\
$^{2}$Okayama Astrophysical Observatory, National Astronomical Observatory of Japan, 3037-5 Honjo, Kamogata, Asakuchi, Okayama 719-0232, Japan\\
$^{3}$Subaru Telescope, National Astronomical Observatory of Japan, 650 North Aohoku Place, Hilo, HI 96720, USA\\
$^{4}$The Graduate University for Advanced Studies, 2-21-1 Osawa, Mitaka, Tokyo 181-8588, Japan\\
$^{5}$Warsaw University Astronomical Observatory, Al. Ujazdowskie 4, 00-478 Warszawa, Poland \\
$^{6}$Instytut Astronomiczny, Uniwersytet Wroc{\l}awski, ul. Kopernika 11, 51-622 Wroc{\l}aw, Poland\\
}

\date{Accepted XXX. Received YYY; in original form ZZZ}

\pubyear{2019}

\begin{document}
\label{firstpage}
\pagerange{\pageref{firstpage}--\pageref{lastpage}}
\maketitle

\begin{abstract}
We present the latest results of our spectroscopic observations and refined modelling
of a sample of
detached eclipsing binaries (DEBs), selected from the \kep Eclipsing Binary
Catalog, that are also double-lined spectroscopic binaries (SB2). 
New high resolution spectra obtained with the HIDES spectrograph, attached 
to the 1.88-m telescope of the Okayama Astrophysical Observatory supplemented the
previous observations, allowing to improve physical parameters of some systems,
and characterize three previously omitted. All the obtained radial velocities (RVs)
were combined with \kep photometry, in order to derive a full set of orbital and 
physical parameters. 

Ten out of eleven SB2s have their component spectra tomographically 
disentangled, and spectral analysis was performed with {\sc ispec}, 
in order to derive effective 
temperatures of components and metallicities of the systems. By comparing
our results with theoretical models, we assess the age and evolutionary
status of the studied objects. We find a good match to all but one systems.
We have derived distances from thus determined parameters, 
and compare them with those from the {\it Gaia} Data Release~2.
For systems previously studied by other authors, our new results
lead to better consistency between observations and models.

%

\end{abstract}




\section{Introduction}

A great source of absolute fundamental parameter of stars are detached 
eclipsing binaries (DEBs) that are also double-lined spectroscopic (SB2). 
They have been used for decades in stellar astrophysics (but not only) for 
many purposes, like testing models of stellar structure and evolution or
for creating empirical relations between various stellar parameters, 
later used as calibrations for studies of single stars, like exoplanet hosts. 
To be useful, the resulting quantities must be known
with sufficient precision \citep[believed to be $\sim$2-3 per cent in masses and 
radii][]{las02,tor10}, the set of presented parameters must be as
complete as possible, and should include individual effective temperatures
of components and, preferably, information about the chemical composition,
such as metallicity and/or abundances of the most important elements (e.g. 
He, Fe, $\alpha$-elements). The precision in absolute physical parameters
can be reached with high-quality radial velocities (RVs) and time series 
photometry. The former can be calculated from high-resolution spectra 
obtained with spectrographs of sufficient stability. The most precise
photometry comes from space-based observatories, such as CoRoT, 
{\it Kepler}, or TESS. The atmospheric parameters usually come from the 
analysis of stellar spectra, provided that the individual spectra are 
properly separated from each other.

All these were the motivation for our program of high-resolution 
spectroscopic monitoring of a sample of bright DEBs from the original 
\kep field, aimed for as precise and complete characterisation of the
observed system, as possible, with the aid of unprecedented precision
of the \kep light curves (LCs). The programme was conducted between July
2014 and November 2017 on the 1.88-m telescope of the Okayama Astrophysical 
Observatory (OAO), equipped with the HIgh-Dispersion Echelle Spectrograph 
\citep[HIDES;][]{izu99}. In this paper we present new
spectroscopic observations, including three new objects, updated
orbital solutions, and, for the first time in this project, spectroscopic
analysis of tomographically disentangled spectra, from which we mainly obtain
the effective temperatures and metallicites, that are later used for 
determination of ages. The manuscript is organized
as follows. In Section~\ref{sec_targets} we present our final sample of
eclipsing SB2s; in Sect.~\ref{sec_data} we present the methodology,
with the focus put on new data and spectroscopic analysis; Sect.~\ref{sec_res}
presents the results, including extended set of parameters, comparison with
literature data and theoretical models that utilizes the newly obtained 
atmospheric parameters, and new hypothesis regarding the tertiary companion 
to an eclipsing SB2; Sect.~\ref{sec_conc} concludes our findings; and finally 
in Sect.\ref{sec_app_k4758} of the Appendix we present updated information 
about one SB1 system from our project, that was also re-observed recently.

\section{The final sample of targets}\label{sec_targets}

\begin{table*}
\centering
\caption{KEBC information about the targets from this work.}\label{tab_cat}
\begin{tabular}{llllllllrcc}
\hline
\hline
KIC & KOI & Other name & RA (deg) & DEC (deg) & $P$~(d)$^a$ & $T_0$ (BJD-2450000)$^a$ & $T_{\rm eff}$ & $kmag$ & $d$ (pc)$^b$ & Stat.$^c$ \\
\hline
\; 3439031 & 4980 & TYC 3134-978-1 & 290.1327 & 38.5137 &  5.9520263 & 4954.068169 & 6337 & 11.287 &  486(6) & N \\
\; 4851217 & 6460 & HD~225524      & 295.8340 & 39.9523 &  2.4702800 & 4953.900507 & 6694 & 11.108 & 1195(53)& N \\
\; 7821010 & 2938 & TYC 3146-1340-1& 291.3199 & 43.5955 & 24.238243  & 4969.615845 & 6298 & 10.816 &  360(5) & RL \\
\; 8552540 & 7054 & V2277~Cyg      & 288.8904 & 44.6170 &  1.0619343 & 4954.105667 & 5749 & 10.292 &  231(1) & --- \\
\; 9246715 & 7601 & HD~190585      & 300.9514 & 45.6041 &171.2776968 & 5170.514777 & 4699 &  9.266 &  616(11)& R \\
\; 9641031 & 7211 & FL Lyr         & 288.0203 & 46.3241 &  2.178154  & 4954.132713 & 5867 &  9.177 & 135.0(5)& RL \\
  10031808 & 7278 & HD~188872      & 298.7976 & 46.9302 &  8.589644  & 4956.430326 &N/A$^d$& 9.557 &  474(6) & --- \\
  10191056~A&5774 & BD+47 2717 A   & 283.8663 & 47.2283 &  2.4274949 & 4955.031469 & 6588 & 10.811 &  613(8) & RL  \\
  10583181 & 7344 & T-Lyr1-01013   & 283.7665 & 47.8190 &  2.6963227 & 4955.210895 & 6231 & 11.009 &  445(5) & N \\ 
  10987439 & 7396 & TYC 3561-922-1 & 296.8259 & 48.4434 & 10.6745992 & 4971.883920 & 6182 & 10.810 &  374(4) & --- \\
  11922782 & 7495 & T-Cyg1-00246   & 296.0074 & 50.2326 &  3.512934  & 4956.247158 & 5581 & 10.460 &  236(2) & --- \\
\hline
\end{tabular}\\
$^a$ For the eclipsing binary, where $T_0$ is the primary eclipse mid-time;
$^b$ From {\it Gaia} DR2 parallaxes \citep{gai16,gai18};\\
$^c$ Status of the system: ``N'' = new, 
not described previously, ``R'' = updated RV solution, 
new spectra presented in this work,\\
``L'' = refined light curve analysis, 
``---'' no new data, same solution as in previous papers, only \isp analysis added; 
\\$^d$ No temperature given in the KEBC.
\end{table*}

In our spectroscopic programme dedicated to bright objects from the \kep 
Eclipsing Binary Catalog 
\citep[KEBC;][]{prs11,sla11,kir16}\footnote{\tt http://keplerebs.villanova.edu/}, 
we observed a total of 22 systems with various characteristics: from blends 
with background stars, through single- and double-lined spectroscopic binaries, 
to multiples containing as much as 5 components. The basic target selection 
criteria were as follows:
\begin{enumerate}
\item \kep magnitude $kmag < 11.5$, to have the targets within the brightness range of 
the telescope. This criterion was initially $kmag < 11.0$, but has changed in 2016, 
when several systems, fainter than $kmag = 11$ were included.
\item Morphology parameter \citep{mat12} $morph < 0.6$ to exclude contact and 
semi-detached configurations.
\item Effective temperature $T_{\rm eff} < 6700$~K to have only late type systems, with many
spectral features. We queried the temperatures from the \kep Input Catalog 
\citep[KIC;][]{kep09}, although some of our objects turned out to be hotter.
\end{enumerate}

Out of the 22 systems observed, to date we published data and analysed 19 of them:
9 single-lined binaries were studied in \citet[][hereafter: Paper~I]{hel16}, 8 double- and
triple-lined systems in \citet[][hereafter: Paper~II]{hel17a}, the double-giant system 
KIC~9246715 was presented in \citet[][hereafter: K924]{hel15}, and a multi-eclipsing quintuple
KIC~4150611 in \citet{hel17a}. In this 
paper we focus on 11 systems, presenting new data for four objects, and three entirely new pairs, 
not yet studied. One of the previously described targets with new observations -- KIC~10191056 -- 
is a triple-lined system, that includes an eclipsing SB2 component (KIC~10191056~A).
This system is the only one in this work for which we did not perform spectral disentangling 
and analysis. Spectral analysis of another triple-lined system -- KIC~6525196 -- is a subject of
another dedicated paper (Alicavus~et~al., in preparation). Flux of the multiple KIC~4150611 
is dominated by a hybrid $\delta$ Scuti/$\gamma$ Doradus pulsator, which itself was also
studied spectroscopically, and his metallicity and age have been estimated \citep{nie15,hel17a}.
Here we also introduce three new SB2s, for which we apply the same working 
approach as to the rest of the sample. 

Below we briefly describe the new objects. Similar descriptions of other targets can
be found in Paper~II and K924.
Unless stated otherwise, this work is the first detailed study for a 
particular object. 
\begin{description}
\item{\it KIC 3439031 = KOI 4980, TYC~3134-978-1}:
This system, shows nearly equal, deep ($\sim$50 per cent) eclipses, suggesting
it is composed of two very similar stars. Except brightness and position 
measurements, no literature data important for this study are available.

\item{\it KIC 4851217 = KOI 6460, HD~225524, HAT~199-10019, ASAS~J194320+3957.1}:
The LC of this system shows strong ellipsoidal variations and, despite the short
orbital period (2.47~d), separation of two minima in phase is different than 0.5, meaning
a non-zero eccentricity. Identified as an eclipsing variable by the HATNET Variability 
Survey \citep{har04}, also listed later in the catalogue of variable stars in the \kep 
field of view of the All Sky Automated Survey \citep[ASAS-K;][]{pig09}.
Its eclipse timing variations (ETV) were studied by \citet{gie12,gie15} and \citet{con14},
and a long-term, parabolic trend in timing of both primary and secondary minima was noted.
Furthermore, \citet{gie12} reported pulsations in the LC, which were later confirmed
to be of a $\delta$ Scuti (dSct) type (Fedurco, priv. comm.). 
Finally, \citet{mat17} collected a number of medium-resolution spectra 
($R\simeq 4000 - 6200$), and derived a preliminary set of orbital and physical parameters
(e.g. $M_1=1.43(5)$ and $M_2=1.55(5)$~M$_\odot$), which we can compare with our results.

\item{\it KIC 10583181 = KOI 7344, T-Lyr1-01013, TYC~3544-2565-1}:
This system has been known as a DEB before the launch of {\it Kepler}. It has been
identified by the Trans-atlantic Exoplanet Survey \citep[TrES;][]{alo04}.
The \kep light curve shows that the secondary eclipse is flat (occultation).
By analysing the TrES data only, \citet{dev08} estimated the masses of both 
components: 1.749(19) and 1.049(15)~M$_\odot$ for the primary and secondary, 
respectively. Our spectroscopy allows us to revise these values. Later, \citet{bor16}
detected strong ETVs with the period of $\sim$3.2~yr,
and derived parameters of the outer orbit. Their solution predicted that the modulation 
of the systemic velocity $\gamma$ of the eclipsing pair will be about 13~k\ms
(peak-to-peak), large enough to be detectable with our spectroscopy.
\end{description}

\section{Data and analysis}\label{sec_data}
We follow the same methodology as in Papers I, II, and K924. They present the observations, 
sources of publicly available data, RV calculations, and approach to RV and LC 
fitting. Here we only describe them briefly, and focus more on those that were not 
previously used.

\subsection{New spectroscopy and RVs}
As in previous works, the HIDES instrument
was fed through a circular fibre, for which the light is collected via a circular 
aperture of projected on-sky diameter of 2.7 seconds of arc, drilled in a flat mirror 
that is used for guiding \citep{kam13}. An image slicer is used in order to reach 
both high resolution ($R\sim50000$) and good efficiency of the system. Spectra 
extraction was done under \textsc{iraf}, using procedures dedicated to HIDES.
Wavelength solution was based on ThAr exposures taken every 1-2 hours, which 
allows for stability of the order of $\sim$40~\ms. The resulting spectra span
from 4360 to 7535~\AA.

The newly presented HIDES observations were done during several runs between
May 2016 and November 2017, with the new observations of objects previously described 
taking place after October 2016. During that time a new queue scheduling mode was introduced 
at the OAO-1.88, and the final observations were done this way, instead of visitor mode.

As in Paper~I, we also made use of the data collected by the Apache Point Observatory Galactic 
Evolution Experiment \citep[APOGEE; ][]{all08,maj17}. We have extracted six individual visit 
spectra\footnote{\tt http://dr12.sdss3.org/advancedIRSearch} of KIC~4851217, but this time we
calculated the RVs ourselves, instead of using the values given by the survey, as in Paper~I. 
Unfortunately, only three of them were recorded when the velocity difference between the
components was large enough to be securely measured. 
Additionally, single archival spectra of KIC~4851217 and KIC~10191056 have been found 
on their ExoFOP-Kepler websites\footnote{\tt https://exofop.ipac.caltech.edu/}. 
These spectra ($R\sim40000$) were taken on July 14, 2014 (KIC~10191056), and 
June 11, 2015 (KIC~4851217) with the TRES spectrograph, 
attached to the 1.5--m Tillinghast telescope of the Fred Lawrence Whipple Observatory 
(FLWO) in Arizona, USA. Two more targets have their TRES spectra available through ExoFOP 
-- KICs~3439031 and 8552540 -- but these observations do not influence the final 
solution significantly, and measured RV agree with models.

Finally, RVs of several of our targets have been reported and analysed by \citet{mat17}.
Apart from the aforementioned KIC~4851217, these are KICs 8552540 and 10191056. For the latter
two objects, the agreement between Paper~II and \citet{mat17} is, generally, very
good. We do not include their RVs in our analysis, as the spectra were taken with
lower resolution, and their inclusion does not change our results significantly.

RV measurements were done with our own implementation of the \textsc{todcor} technique 
\citep{zuc94}, which finds velocities of two stars $v_1$ and $v_2$ simultaneously. 
As templates for the HIDES and TRES data we used synthetic spectra computed with 
ATLAS9 and ATLAS12 codes \citep{kur92}, which do not reach wavelengths 
longer than 6500~\AA. For the APOGEE observations we used synthetic spectra
from the library of \citet{coe05}, which cover both optical and IR regions
(3800-18000~\AA), but have lower resolution than the ones from ATLAS9/12. 
Single measurement errors were calculated with a bootstrap approach \citep{hel12}, and used for 
weighting the measurements during the orbital fit, as they are sensitive to 
the signal-to-noise ratio (SNR) of the spectra and rotational broadening of the 
lines. All (previous and new) individual RV measurements are presented in 
Table~\ref{tab_RV} in the Appendix.

\subsection{\kep photometry}
The RVs were supplemented by the long- and short-cadence \kep photometry, which
is available for download from the KEBC. We used the de-trended relative flux measurements $f_{\rm dtr}$, 
that were later transformed into magnitude difference $\Delta m=-2.5 \log(f_{\rm dtr})$,
and finally the catalogue value of $kmag$ was added. From the systems presented here,
only KICs 9246715 and 10987439 do not have short-cadence data available.

Even though our new systems have time-series photometry available from other sources,
like the TrES survey, we only work on the \kep data. Comparison of solutions based on
\kep observations with ones that base on TrES or ASAS data, has been discussed in Paper~II.

\subsection{Orbital solutions}
The orbital solutions were found using our own procedure called {\sc v2fit}
\citep{kon10}. We used it mainly to fit a double-Keplerian orbit to a set
of RV measurements of two components, utilizing the Levenberg-Marquardt
minimization scheme. The fitted parameters are: orbital period $P$, zero-phase 
$T_P$\footnote{Defined in this code as the moment of passing the pericentre for eccentric 
orbits or quadrature for circular.}, systemic velocity $\gamma$, velocity 
semi-amplitudes $K_{1,2}$, eccentricity $e$ and periastron longitude $\omega$.
Depending on the case,
we also included such effects as: the difference between systemic velocities of 
two components, $\gamma_2-\gamma_1$, linear and quadratic trends, or periodic
modulation of $\gamma$ caused by a circumbinary body on an outer orbit, parametrized 
analogously by orbital parameters $P_3$, $T_3$, $K_3$, $e_3$, and $\omega_3$. 
In such case $\gamma$ is defined in the code as the systemic velocity of the  whole triple. 
Whenever applicable, we simplified our fit by keeping the orbital period on the 
value given in the KEBC. Whenever $\gamma_2-\gamma_1$ or $e$ were found
to be not significantly different from zero, the fit was repeated 
with those parameters fixed to 0. 
For KIC~4851217 we also searched for the difference between zero points of 
HIDES and APOGEE. This was not done for HIDES vs. TRES zero-points, because 
only one TRES spectrum per target is available.

Systematics that come from fixing a certain parameter in the fit are assessed 
by a Monte-Carlo procedure, and other possible systematics (like coming from 
poor sampling, low number of measurements, pulsations, etc.) by a bootstrap 
analysis. All the uncertainties of orbital parameters given in this work 
already include the systematics.

Moreover, to obtain reliable formal parameter errors of the fit, and the final,
reduced $\chi^2$ to be close to 1, we were modifying the RV measurement errors 
either by adding a systematic term (jitter) in quadrature, or multiplying by a 
certain factor. Adding the jitter works better for active stars, when the RV
scatter is caused by spots, and is compensated with the additional term. However,
since \vfit weights the measurements on the basis of their own errors, 
which are sensitive to SNR and rotational velocity, we mainly used the 
second option, in which the weights are preserved.

\subsection{Light curve solutions}\label{sec_jkt}

The \kep light curves were fitted with the version 28 (v28) 
of the code \jkt \citep{sou04a,sou04b}, which is based on the 
\textsc{ebop} program \citep{pop81}. 

The short-cadence data, due to their
denser time sampling, may include information that is missing in the long cadence,
e.g. better reperesent short-time-scale brightness variations, like egress and ingress 
of some eclipses. We have therefore fitted both kinds of \kep photometry, taking into
account availability of the short-cadence data, and compared the output. 

For long-cadence data, the errors were estimated with a residual-shift (RS) method 
\citep{sou11}, run on data from each quarter separately, as described in Paper~I. 
This was done in order to properly account for strong systematic effects of different time 
scales (short-term pulsations or long-term evolution of spots). For short-cadence 
data, the same approach would take months of computer time, so instead of RS we used the 
available Monte-Carlo (MC) option, also run on each quarter separately. In both cases,
to get the the final uncertainties, we added in quadrature the formal error
of a weighted average of all available quarters, and the $rms$ of the results from each
quarter or set (Paper~I).

Starting values of eccentricity $e$ and periastron longitude $\omega$,
as well as mass ratio $q$ (here held fixed), were taken from \vfit runs.
We fitted for the period $P$, mid-time of the 
primary (deeper) minimum $T_0$, sum of the fractional radii $r_1 + r_2$ 
(where $r = R/a$), their ratio $k$, inclination $i$, surface brightness 
ratio $J$, maximum brightness $S$, as well 
as for $e$ and $\omega$ (their final values are from the \jkt runs,
unless stated otherwise). Third light contribution $l_3/l_{\rm tot}$,
which can be significant in \kep data, was also initially fitted for,
but when it was found not significantly different from zero, or even negative,
the fit was repeated with fixed $l_3/l_{\rm tot}=0$.
 The gravity darkening coefficients were always kept 
fixed at the values appropriate for stars with convective envelopes ($g = 0.32$).
We did not fit for limb darkening (LD) coefficients, but we found them iteratively,
as described in Paper~II, and perturbed them in the RS or MC stage.

The final values of $P$ and $T_0$ were derived from the complete long-cadence curves.
However, since various systems show different LC characteristics (width of eclipses, ellipsoidal
variations, pulsations, evolving spots, flares, timing variations, etc.), 
each binary has been treated individually. In Subsections describing results obtained for 
each target, we explicitly state if the adopted results come from short- or long-cadence
LCs, if from fit to a complete curve, or the weighted average from single-quarter curves, 
and which approach (RS or MC combined with $rms$) has been used to obtain uncertainties.

\subsection{Tomographic spectra disentangling}

\begin{table*}
\centering
\caption{Summary of TD information. Number of input spectra ($N$ sp.), final wavelength range, and SNRs are given.}\label{tab_td_spec}
\begin{tabular}{lrcrr}
\hline
\hline
KIC & $N$ sp. & Wavelength ranges [\AA] & SNR$_1$ & SNR$_2$ \\
\hline
\; 3439031 &  8 & 5030.555 -- 6222.286 &  70 &  69 \\
\; 4851217 &  8 & 5030.316 -- 6221.982 &  47 & 115 \\
\; 7821010 & 15 & 5265.545 -- 6222.346 & 129 &  94 \\
\; 8552540 &  8 & 5030.468 -- 6222.174 & 136 &  37 \\
\; 9246715 & 17 & 5030.582 -- 5204.669; 5216.799 -- 5400.960; 5417.327 -- 5901.768; 5930.232 -- 6222.319 & 123 & 117 \\
\; 9641031 & 14 & 5030.575 -- 5400.956; 5417.321 -- 6155.450 & 272 & 64 \\
  10031808 & 16 & 5030.582 -- 5400.958; 5417.326 -- 5841.582; 5930.227 -- 6222.319 & 136 & 170 \\
  10583181 &  9 & 5030.323 -- 6221.987 & 146 & 14 \\
  10987439 & 10 & 5469.778 -- 5841.463; 5868.371 -- 6222.195 &  20 & 134 \\
  11922782 & 10 & 5030.582 -- 5204.667; 5216.978 -- 6222.319 & 149 &  22 \\
\hline
\end{tabular}
\end{table*}

In order to obtain separate spectra for each component, we performed 
tomographic disentangling (TD) of composite spectra of 10 systems.
This was done when at least 8 good quality observations were available, 
and only two sets of lines were clearly seen. For this reason we omitted 
the triple-lined KIC~10191056, but included double-lined KIC~10583181, 
where significant third light was noted in the \kep LC (Section~\ref{sec_res}).

We used the method described in \citet{kon10}, which is based on tomographic 
approach proposed by \citet{bag91}. It utilizes prior RV measurements, which were
made with {\sc todcor}. The algorithm works on each echelle order 
separately, and is fragile to low-signal data. For this reason the edges of each 
order were first trimmed, not every order was included or produced a 
satisfactory solution, and, in particular, we only used the orders from the ``green''
chip of the HIDES detector\footnote{\tt 
http://www.oao.nao.ac.jp/$\sim$hides/wiki/index.php\\ ?Mosaic\_CCD\_en}. 
Therefore, the wavelength range of final spectra varies from target to target. 
In a low number of cases, a given low-SNR spectrum was not used for TD, but still
gave reasonably precise RV measurements.
Before the TD, each order has been continuum-normalized. In the final steps, we merged 
the orders into single, long, 1-dimensional spectra.
In Table~\ref{tab_td_spec} we show for each star how many 
input spectra were used, what was the final wavelength range (inc. gaps), and
what were the SNRs of disentangled primary and secondary spectra. Since the primary 
is defined by the LC, it is not always the brighter component. 

We should note that two HIDES orders (96 and 97) containing the sodium D lines 
(around 5890~\AA) did, formally, give a tomographic solution. However, the contamination
from the interstellar Na introduced a third set of strong lines, therefore
pieces of final spectra around the D lines are not recovered correctly and were
not used in further analysis. 

Before further analysis, the resulting TD spectra needed to be renormalized, 
in order to obtain the true intensities of the spectral lines. This can be done with
the aid of additional information about the ratio of fluxes at a given wavelength
(here: echelle order), which can be taken from {\sc todcor}. The value of flux 
ratio $\alpha$ that maximizes the height of the cross-correlation
can be calculated with a relatively simple analytical relation \citep[equation 
A4 in][]{zuc94}. The usage of \todcor flux ratios has been shown to give proper 
results for example in \citet{hel15b} or \citet{bri17}.

\subsection{Spectroscopic analysis}

The main motivation for this work was to complete the set of stellar parameters
of the studied DEBs with effective temperatures and metallicities, which are necessary 
for further determination of the age and evolutionary status. For this we used the
v2018.06.08 version of the freely distributed code \isp \citep{bla14}, and our TD spectra.
Each spectrum was first corrected for a residual RV shift, introduced in the TD stage 
(typically 0.3-0.5 k\ms), and resampled with the wavelength step of 0.05~\AA. The TD 
output is severely oversampled, so this step allowed to reduce the number of data points
in each spectrum, making the analysis about 10$\times$ faster, while keeping the original
HIDES resolution ($R\sim50000$). To ensure the uncerainties given by \isp are
trustworthy, we used the program to calculate reliable flux errors, basing on 
the measured SNR of the spectra. With these errors the reduced $\chi^2$ of the fit was
tipically close to 1, with the exception of low-SNR spectra ($<$50), which analysis we do not
find reliable.

To find the atmospheric parameters we used the spectral synthesis approach, utilising 
the code {\sc spectrum} \citep{gra94}, the MARCS grid of model atmospheres \citep{gus08}, 
and solar abundances from \citet{gre07}. \isp synthesizes spectra only in certain,
user-defined ranges, called ``segments''. We followed the default approach, where these 
segments are defined as regions $\pm$2.5~\AA\,around a certain line. We decided to synthesize
spectra around a set of lines carefully selected by the the {\it Gaia}-ESO Survey 
\citep[GES; ][]{gil12,ran13} in such way, that various spectral fitting codes reproduce 
consistent parameters from a reference solar spectrum \citep{bla16}.

With several exceptions, described below, we run the fit with the following parameters
set free: effective temperature $T_{\rm eff}$, metallicity [M/H], alpha enhancement 
[$\alpha$/Fe], and microturbulence velocity $v_{\rm mic}$. The resolution $R$ was
always fixed to 50000, and gravity $\log(g)$ to the value corresponding to absolute values 
of mass and radius (see next Section), which is more precise than $\log(g)$ we could find from 
spectroscopy. In case of short-period, circular (or nearly circular) orbits, where stars
rotate (pseudo-)synchronously with the orbital period, the rotational velocity $v\sin(i)$ 
was also fixed and set to values expected from the synchronous rotation. In only three
cases we set $v\sin(i)$ free: KIC~7821010, 9246710 and 10031808. In the first two 
the lines are quite narrow, so $v_{\rm mic}$ was automatically calculated by \isp
from an empirical relation found by the GES Consortium 
and incorporated into the \isp program (GES Consortium, priv. communication). Lines of 
KIC~10031808 are rotationally broadened and independent fit for $v_{\rm mic}$ was possible.
The macroturbulence velocity $v_{\rm mac}$, which degenerates with rotation, was at all 
times calculated on-the-fly by \isp from a similar empirical relation found by the GES
Consortium. 

As final values of systemic [M/H] and [$\alpha$/Fe] we adopted averages of values obtained
from each component. As their conservative uncertainties we added in quadrature the formal 
parameter errors from \isp and half the difference between two individual results. 
For example, for KIC~10031808 from [M/H]$_1=-0.17(5)$ and [M/H]$_2=-0.05(5)$, we 
got [M/H]$=-0.11(8)$. This particular example, and the metallicity of KIC~9641031
([M/H]$_1=-0.15(6)$, [M/H]$_2=0.00(6)$, adopted [M/H]$=-0.07(9)$) are the only two
where individual values differ by more than formal 1$\sigma$, but still within 2$\sigma$.
When only one component could be analysed, we adopted its [M/H] and 
[$\alpha$/Fe] for the whole system. Abundances of specific elements were not calculated.


\subsection{Calculation of absolute parameters}

The partial results of LC and RV solutions were later combined in order to calculate 
the absolute values of stellar parameters using the \jktabs code, available 
together with the {\sc jktebop}. As an input, this simple procedure takes orbital 
period, eccentricity, fractional radii, velocity semi-amplitudes and inclination 
(all with uncertainties), and returns absolute values of masses and radii (in solar 
units), $\log(g)$ and rotational velocities, assuming tidal locking and 
synchronization. 

\jktabs can also calculate distance to an object, using effective 
temperatures of two components, approximate metallicity (given with precision of 0.5 dex), 
$E(B-V)$ and apparent magnitudes. The code does not work on brightnesses in \kep band, 
so, unless stated otherwise, for the distance estimation we used $B,V,J,H$ and $K$-band entries 
from {\it Simbad} \citep{wen00}. We only used the temperatures and [M/H] from the \isp analysis.
As the final value of distance we adopt a weighted average of five values,
calculated for each band from the surface brightness-$T_{\rm eff}$ relations of 
\citet{ker04}. The results can be compared with parallaxes from the
{\it Gaia} Data Release~2 \citep[GDR2; ][]{gai16,gai18}.

\begin{landscape}
\begin{table}
\scriptsize
\caption{Orbital and physical parameters of eleven double-lined eclipsing
binaries from our survey, including results of spectral analysis, when possible. 
When ``a'' is given in parenthesis instead of uncertainty, the parameter was 
calculated automatically from an empirical relation.}\label{tab_par_sb2}
\begin{tabular}{lccccccccccc}
\hline \hline
KIC                     & {\bf 3439031} &  4851217      & {\bf 7821010} & 8552540       & 9246715       & {\bf 9641031} & 10031808      & {\bf 10191056~A} & {\bf 10583181}& 10987439      & 11922782  \\
\hline
$P_{ecl}$ (d)           & 5.95202634(74)& 2.4702836(17) &  24.238235(4) & 1.06193441(4) &   171.2770(6) & 2.17815425(7) & 8.5896432(13) &2.427494881(19)& 2.696357(41)  &10.67459809(33)&  3.5129340(3) \\
$T_0$ (JD-2454900)$^a$  & 57.040009(11) & 53.899973(53) &  69.61678(13) &  54.105945(27)&   99.2536(31) &  54.133349(3) &  56.43099(10) &  55.031699(5) &  55.210(13)   &  71.885044(32)&  56.24790(7)  \\
$T_P$ (JD-2454900)$^b$  &  56.34(79)    &  54.66(27)    &  69.341(22)   &  53.846(23)   &   81.780(45)  &  53.58761(56) &  56.475(74)   &  53.9355(11)  &  54.5303(34)  &  60.136(72)   &  51.856(20)   \\
$K_1$ (k\ms)            &  79.171(67)   & 131.0(2.6)    &  66.50(35)    & 121.0(1.6)    &   33.213(21)  &  93.34(10)    &  83.08(28)    & 106.96(58)    &  79.33(50)    &  76.39(10)    &  76.04(29)    \\
$K_2$ (k\ms)            &  79.274(84)   & 114.6(2.7)    &  69.55(36)    & 145.9(2.0)    &   33.629(22)  & 118.75(35)    &  80.42(15)    & 118.34(33)    & 128.8(1.3)    &  53.00(9)     &  97.01(42)    \\
$\gamma_1$ (k\ms)       &  27.023(52)   & -22.9(2.2)    & -17.15(46)    & -14.1(1.1)    &   -4.532(16)  & -37.38(8)     &  12.96(17)    & -25.68(23)    & -40.2(1.2)    & -19.11(11)    & -41.84(13)    \\
$\gamma_2-\gamma_1$ (k\ms)& 0.0(fix)    &   0.0(fix)    &   0.0(fix)    &   0.0(fix)    &   -0.055(22)  &   0.0(fix)    &  -0.17(21)    &   0.0(fix)    &   2.3(6)      &  -0.37(19)    &   0.0(fix)    \\
$q$                     &   1.0013(14)  &   1.143(35)   &   0.956(7)    &   0.829(16)   &    0.9876(9)  &   0.7860(23)  &   1.033(4)    &   0.9038(55)  &   0.6160(74)  &   1.4413(31)  &   0.784(4)    \\
$M_1\sin^3(i)$ (M$_\odot$)& 1.2258(24)  &   1.769(88)   &   1.276(17)   &   1.144(36)   &    2.1782(33) &   1.2056(72)  &   1.705(9)    &   1.511(12)   &   1.558(36)   &   0.9776(34)  &   1.057(10)   \\
$M_2\sin^3(i)$ (M$_\odot$)& 1.2274(28)  &   2.021(94)   &   1.220(16)   &   0.948(28)   &    2.1512(32) &   0.9477(35)  &   1.762(13)   &   1.365(15)   &   0.960(17)   &   1.4090(45)  &   0.829(6)    \\
$a\sin(i)$ (R$_\odot$)  &  18.646(13)   &  11.99(18)    &  47.83(20)    &   5.604(54)   &  211.59(10)   &   9.134(15)   &  26.723(54)   &  10.813(32)   &  11.095(75)   &  27.272(28)   &  12.019(32)   \\
$e$                     &   0.00142(16) &   0.0311(7)   &   0.6796(16)  &   0.0(fix)    &   0.3552(4)   &   0.0(fix)    &   0.2717(14)  &   0.00283(23) &   0.0(fix)    &   0.0509(14)  &   0.0(fix)    \\
$\omega$($^\circ$)      &   47(48)      & 188(10)       &  59.47(22)    &   ---         &  18.22(11)    &   ---         &  94.067(70)   & 284(4)        &   ---         &  51.7(2.3)    &   ---         \\
$r_1$                   &   0.07547(10) &   0.1659(64)  &  0.02669(20)  &   0.2509(31)  &   0.04008(58) &   0.1361(25)  &   0.09628(71) &   0.1807(19)  &   0.1696(6)   &   0.03409(54) &   0.1245(47)  \\
$r_2$                   &   0.07524(13) &   0.2607(32)  &  0.02530(27)  &   0.1806(40)  &   0.03870(40) &   0.0984(26)  &   0.11250(47) &   0.1517(27)  &   0.0883(7)   &   0.0553(11)  &   0.0704(52)  \\
$i$ ($^\circ$)          &  89.57(1)     &  77.11(26)    &  89.58(2)     &  85.83(46)    &  87.049(31)   &  87.13(71)    &  83.323(47)   &  81.33(8)     &  89.4(8)      &  85.614(66)   &  85.52(60)    \\
$J$                     &   0.9970(26)  &   0.932(43)   &   0.922(23)   &   0.67(14)    &   1.042(49)   &   0.396(49)   &   0.903(34)   &   0.945(9)    &   0.411(14)   &   2.56(22)    &   0.46(11)    \\
$l_2/l_1$               &   0.9933(51)  &   2.32(23)    &   0.83(4)     &   0.292(10)   &   0.964(48)   &   0.224(35)   &   1.2303(23)  &   0.68(4)     &   0.105(3)    &   6.48(1.14)  &   0.15(3)     \\
$l_3/l_{\rm tot}$       &$\sim$0.0(var) &   0.0(fix)    &   0.0(fix)    &   0.0(fix)    &   0.0(fix)    &   0.0(fix)    &   0.0(fix)    &   0.160(9)    &$\sim$0.12(var)&   0.0(fix)    &   0.0(fix)    \\
$M_1$ (M$_\odot$)       &   1.2259(24)  &   1.91(10)    &   1.277(17)   &   1.153(36)   &   2.1869(33)  &   1.2102(76)  &   1.741(9)    &   1.564(12)   &   1.559(36)   &   0.9862(34)  &   1.067(10)   \\
$M_2$ (M$_\odot$)       &   1.2275(28)  &   2.18(10)    &   1.221(16)   &   0.956(28)   &   2.1598(32)  &   0.9512(39)  &   1.798(13)   &   1.413(16)   &   0.960(16)   &   1.4215(45)  &   0.836(6)    \\
$R_1$ (R$_\odot$)       &   1.4072(22)  &   2.041(85)   &   1.276(11)   &   1.410(22)   &   8.49(12)    &   1.244(23)   &   2.590(20)   &   1.976(22)   &   1.882(14)   &   0.932(15)   &   1.501(57)   \\
$R_2$ (R$_\odot$)       &   1.4030(25)  &   3.207(63)   &   1.210(14)   &   1.015(24)   &   8.20(9)     &   0.900(24)   &   3.027(14)   &   1.659(30)   &   0.980(10)   &   1.512(31)   &   0.849(63)   \\
$a$   (R$_\odot$)       &  18.646(13)   &  12.30(19)    &  47.83(20)    &   5.619(54)   & 211.87(10)    &   9.145(16)   &  26.905(54)   &  10.938(33)   &  11.095(75)   &  27.352(29)   &  12.056(34)   \\
$\log(g_1)$             &   4.2300(13)  &   4.100(35)   &   4.332(7)    &   4.202(12)   &   2.920(13)   &   4.331(16)   &   3.852(6)    &   4.041(9)    &   4.082(5)    &   4.493(14)   &   4.114(33)   \\
$\log(g_2)$             &   4.2332(15)  &   3.765(13)   &   4.359(9)    &   4.406(20)   &   2.945(9)    &   4.508(23)   &   3.731(4)    &   4.149(16)   &   4.438(7)    &   4.232(18)   &   4.503(64)   \\
$T_{\rm eff,1}$ (K)     &   6530(220)   &   ---         &   6720(160)   &   6060(200)   &   4890(50)    &   6260(120)   &   7105(110)   &   ---         &   6730(140)   &   ---         &   5990(110)   \\
$T_{\rm eff,2}$ (K)     &   6530(220)   &   7250(215)   &   6570(190)   &   ---         &   4905(60)    &   5490(240)   &   6840(105)   &   ---         &   ---         &   6490(90)    &   ---         \\
$\log(L_1)$ (L$_\odot$) &   0.45(6)     &   ---         &   0.47(4)     &   0.381(59)   &   1.567(22)   &   0.330(35)   &   1.185(27)   &   ---         &   0.824(37)   &   ---         &   0.416(45)   \\
$\log(L_2)$ (L$_\odot$) &   0.45(6)     &   1.406(54)   &   0.39(5)     &   ---         &   1.543(22)   &  -0.180(80)   &   1.254(27)   &   ---         &   ---         &   0.550(30)   &   ---         \\
$v_{\rm rot,1}$	(k\ms)$^c$&11.96(2)s    &  41.8(1.7)s   &   1.3(4.8)    &  67.1(1.0)s   &   2.56(22)    &  28.89(54)s   &  14.9(1.1)    &  41.2(5)s     &  35.31(26)s   &   4.42(7)s    &  21.61(82)s   \\
$v_{\rm rot,2}$	(k\ms)$^c$&11.92(2)s    &  65.7(1.3)s   &   8.9(1.1)    &  48.3(1.2)s   &   2.63(22)    &  20.89(55)s   &  10.2(2.0)    &  34.6(6)s     &  18.39(19)s   &   7.16(15)s   &  12.22(90)s   \\
$v_{\rm mic,1}$	(k\ms)  &   1.63(21)    &   ---         &   1.54(a)     &   1.28(40)    &   1.27(a)     &   1.90(18)    &   2.76(25)    &   ---         &   2.20(27)    &   ---         &   1.63(18)    \\
$v_{\rm mic,2}$	(k\ms)  &   1.72(21)    &   1.70(47)    &   1.44(a)     &   ---         &   1.27(a)     &   1.79(54)    &   3.26(31)    &   ---         &   ---         &   1.88(21)    &   ---         \\
$v_{\rm mac,1}$	(k\ms)  &   9.24(a)     &   ---         &  11.13(a)     &   5.99(a)     &   3.83(a)     &   7.23(a)     &  13.01(a)     &   ---         &  11.46(a)     &   ---         &   5.57(a)     \\
$v_{\rm mac,2}$	(k\ms)  &   9.29(a)     &  18.4(a)      &   9.56(a)     &   ---         &   3.77(a)     &   3.33(a)     &  16.21(a)     &   ---         &   ---         &   8.97(a)     &   ---         \\
$[$M/H$]$               &   0.10(13)    &  -0.08(9)     &   0.10(8)     &  -0.27(11)    &   0.01(3)     &  -0.07(9)     &  -0.11(8)     &   ---         &  -0.01(7)     &  -0.03(5)     &  -0.22(6)     \\
$[\alpha$/Fe$]$         &   0.03(14)    &  -0.03(15)    &   0.02(8)     &  -0.01(11)    &  -0.01(3)     &   0.03(9)     &   0.16(6)     &   ---         &   0.00(7)     &   0.05(16)    &  -0.01(6)     \\
$d_{\rm J}$ (pc)$^d$    &   477(15)     &   ---         &   342(13)     &   ---         &   573(13)     &   130(5)      &   445(15)     &   ---         &   ---         &   ---         &   ---         \\
$rms_{\rm RV1}$ (k\ms)  &   0.15        &   3.80        &   0.092       &   3.02        &   0.046       &   0.39        &   0.50        &   1.70        &   0.70        &   0.129       &   0.37        \\
$rms_{\rm RV2}$ (k\ms)  &   0.17        &   4.82        &   0.217       &   5.26        &   0.051       &   0.83        &   0.30        &   1.08        &   2.50        &   0.042       &   1.02        \\
$rms_{\rm LC}$ (mmag)$^e$&  0.36        &   4.25        &   0.50        &  12.7         &   0.61        &   5.44        &   1.22        &   0.41        &   2.06        &   0.072       &   8.06        \\
$rms_{\rm SC}$ (mmag)$^e$&  0.58        &   5.13        &   0.45        &  14.6         &   ---         &   5.50        &   1.26        &   0.40        &   1.69        &   ---         &   5.60        \\
\hline 
\end{tabular}\\
$^a$ Mid-time of the primary (deeper) eclipse, calculated from the complete Q0-Q17 curve.\\
$^b$ Time of pericentre or quadrature.\\
$^c$ If ``s'' is given after the uncertainty, it is the velocity of (pseudo-)synchronous rotation given by {\sc jktabsdim}. Otherwise, it is $v\sin(i)$ calculated by {\sc ispec}.\\
$^d$ The \jktabs distance $d_{\rm J}$ is calculated only when both $T_{\rm eff}$-s were found with {\sc ispec}.\\
$^e$ {\bf The $rms$ of residuals of the fit to the complete long- (LC) and short-cadence (SC) \kep curve.}
\end{table}
\end{landscape}

\subsection{Comparison with isochrones and age estimation}\label{sec_iso_met}

We use our results to assess the age $\tau$ and evolutionary 
status of each system. We compare them to isochrones generated with
Modules for Experiments in Stellar Astrophysics (MESA) with the aid of
MESA Isochrones \& Stellar Tracks (MIST v1.2) web 
interface\footnote{\tt http://waps.cfa.harvard.edu/MIST/} \citep{pax11,cho16,dot16}.
To assess the age $\tau$ of a system, we searched for the best-fitting isochrone 
simultaneously on the $M/R$ plane for both components, and $M/T_{\rm eff}$ 
for one or two stars, depending on how many $T_{\rm eff}$s are known from 
spectroscopy. We adopt metallicities from the \isp analysis, and 
followed the assumption used in MIST that [Fe/H] = [$Z$/H]. In current version
MIST assumes protosolar abundance $Z=0.0142$ from \citet{asp09}.

\section{Results}\label{sec_res}

\subsection{Parameters of eclipsing binaries}\label{sec_par}

In this section we present results of the combined {\sc v2fit + jktebop + ispec}
analysis of ten double-lined eclipsing binaries, and updated {\sc v2fit + jktebop}
of KIC~10191056~A, which is the close pair in a triple-lined hierarchical
system. They can be found in Table~\ref{tab_par_sb2}. Observed and modelled RV 
and LC curves are presented in Figures \ref{fig_mod_0343} to \ref{fig_mod_1192}.

\subsubsection{KIC~3439031}

\begin{figure*}
\includegraphics[width=0.9\columnwidth]{K0343_orb.eps}
\includegraphics[width=0.9\columnwidth]{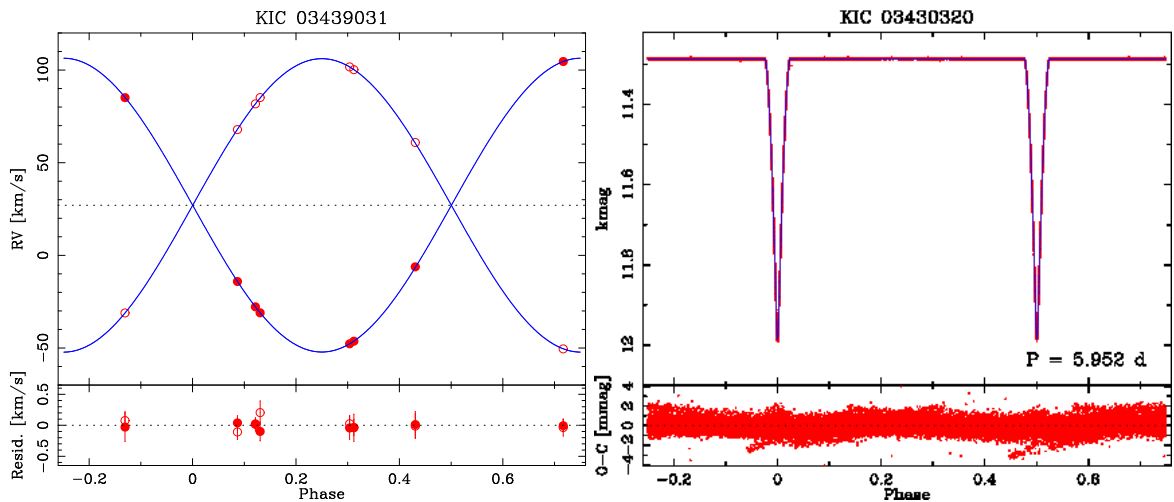}
\caption{Radial velocity (left) and light (right) curves of KIC~3439031. 
The best-fitting models are plotted with blue lines. Filled circles on the 
RV plot refer to HIDES data for the primary, and open ones for the secondary. 
On both sides the phase 0 is set for the deeper eclipse mid-time, 
according to the definition in \textsc{jktebop}.
}\label{fig_mod_0343}
\end{figure*}

This system has been observed with HIDES 8 times, and was not studied so far.
It is the faintest target in our sample.
We found it is composed of two nearly identical F-type stars. 
Despite the number of RV measurements is low, the $rms$ of 
the fit is very good, hence the mass uncertainty is low as well: $\sim$0.2~per cent
(including systematics coming from the low number of data points). 

This system has the \kep short-cadence photometry available for parts of quarters
Q02 and Q04, and for entire quarters Q11 and Q12. Eclipses in the short-cadence 
curve are deeper and last shorter, meaning that the long-cadence data suffered from 
the longer exposure time and averaging the brightness variability. However, this 
system is quite stable out of eclipses, and we decided that the short-cadence data 
(presented in Figure~\ref{fig_mod_0343}) are sufficient to
analyse it properly. Furthermore, in both short- and long-cadence data we noticed that 
the residuals in the eclipses behaved differently from quarter-to-quarter, e.g. the 
model eclipses were too shallow in long-cadence quarters Q04 and Q05, then suddenly 
change to too deep in Q06 and Q07. We suspect that it is a systematic related to the 
rotation and orientation of the telescope. To compensate for it, we decided to analyse 
the LC of KIC~3439031 quarter-by-quarter and adopt weighted averages as final values of 
parameters. The uncertainties were calculated with the previously described
procedure that combines $rms$ of results of single-quarter fits with individual errors 
from a Monte-Carlo analysis.
 
Apart from eclipses and ellipsoidal variations, the LC shows little variability.
The $rms$ of the fit is therefore quite low (0.33 and 0.58~mmag for long- and
short-cadence curves, respectively), hampered mainly by the systematics coming
from imperfect de-trending and telescope's pointing stability. Final precision in radii is at 
a very good level of $\sim$0.15-0.18~per cent. The \isp analysis gave similar 
results for both components, which was expected for a pair of nearly identical stars. 
The system appears to be slightly more metal rich than the Sun, and $\alpha$ elements 
are not significantly enhanced. Stars are rotating synchronously with the orbital 
period, and the orbit is nearly circular. The time scale of circularisation (as given by
{\sc jktabsdim}) is 4.27~Gyr, which may be treated as an upper limit of the 
system's age. Effective temperatures are representative of the F6 spectral type,
and are about 200~K higher than estimates from KEBC or GDR2, but this is within the
error bars of our values. Disentangled spectra of higher SNR are needed to
lower the uncertainties. When used in {\sc jktabsdim}, the $T_{\rm eff}$s give the 
distance $d_{\rm J} = 477(15)$~pc, which is in a reasonable agreement with 
486(6)~pc from GDR2. The distance calculation assumes no reddening, as the
consistency between distances derived for different bands was already quite good.

The set and precision of parameters we provide for KIC~3439031, 
and the fact that it consist a pair of ``twins'', make this target
valuable for testing stellar formation and evolution models.

\subsubsection{KIC~4851217}

\begin{figure*}
\includegraphics[width=0.9\columnwidth]{K0485_orb.eps}
\includegraphics[width=0.9\columnwidth]{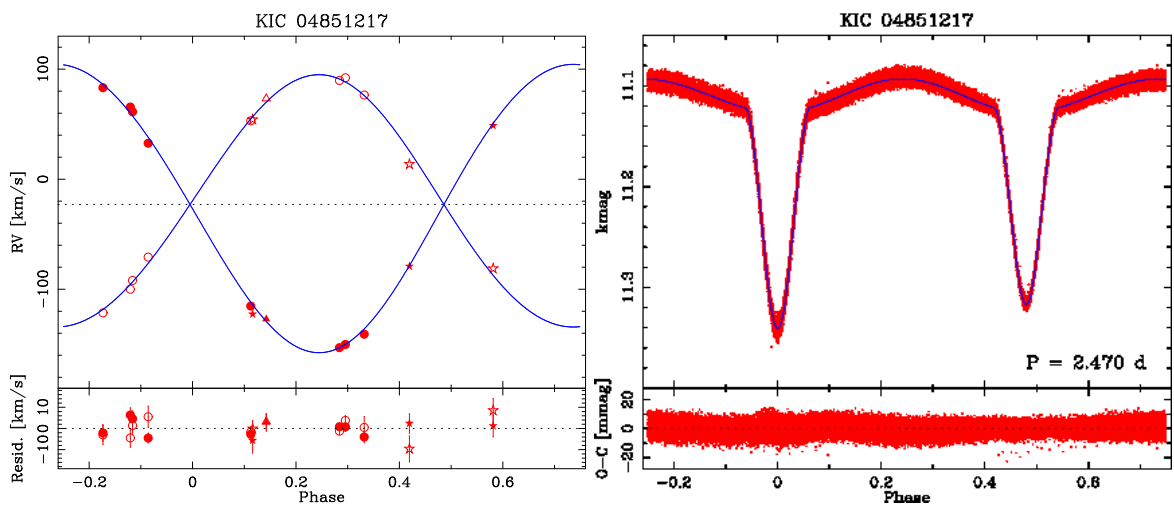}
\caption{Same as Fig. \ref{fig_mod_0343}, but for KIC~4851217 and with
additional RV measurements from APOGEE (stars) and TRES (triangles).
}\label{fig_mod_0485}
\end{figure*}

\begin{table}
\centering
\caption{Comparison of updated results for KIC~4851217 with parameters from 
\citet{mat17}.}\label{tab_kic485}
\begin{tabular}{lcc}
\hline \hline
	&This work	& \citeauthor{mat17}\\
Parameter & (Table \ref{tab_par_sb2}) & (\citeyear{mat17})\\
\hline
$P$ (d)		&     2.4702836(17)	& 2.47028283(-)	\\
$K_1$ (k\ms)	& 131.0(2.6)	& 115(2)	\\
$K_2$ (k\ms)	& 114.6(2.7)	& 107(2)	\\
$q$				&  1.143(35)	& 1.08(3)	\\
$i$ ($^\circ$)	&  77.11(26)	&  77.8(-)	\\
$a$ (R$_\odot$)	&  12.30(19)	& 11.1(2)	\\
$M_1$ (M$_\odot$)	& 1.19(10) & 1.43(5)	\\
$M_2$ (M$_\odot$)	& 2.18(10) & 1.55(5)	\\
$rms_1$ (k\ms)	& 3.8	& 7.0	\\
$rms_2$ (k\ms)	& 4.8	& 7.2	\\
\hline
\end{tabular}
\end{table}

Comparison of \jkt output for short- and long-cadence data showed that
the two LCs are indistinguishable in shape, and the resulting parameters
differ typically by less than 1/5 of the obtained errors. We therefore adopted 
the result of a fit to the complete Q00-Q17 long-cadence curve as the final ones.
The long-cadence uncertainties were slightly higher (by roughly 10-20 per cent) 
from those from the short-cadence curve, and we decided to use them, as they are 
the more conservative ones, probably better coping with the influence of pulsations.
Figure~\ref{fig_mod_0485} shows the long-cadence curve.

We observed this system 8 times with HIDES, and additionally used three
APOGEE and one TRES spectra. We did this to double-check and strengthen
our orbital solution (Fig.~\ref{fig_mod_0485}), which initially was in 
disagreement with that from \citet{mat17}. Our data clustered around phase 
$\phi\sim 0.3$ and between $\phi$ = 0.8 -- 1.0. With the additional spectra, 
our sampling is more uniform, and the velocity amplitudes ($K_1=131\pm3$, 
$K_2=115\pm3$~k\ms; Table~\ref{tab_par_sb2}) are much better constrained. The original solution from 
\citeauthor{mat17}, in which $K_1=115\pm2$, $K_2=107\pm2$~k\ms, suffers from a 
similar problem (data only for $\phi =$ 0.25 -- 0.5 and 0.8 -- 0.1), which may
explain the disagreement. A full comparison is shown in Table~\ref{tab_kic485}.
A recent re-analysis of these data results in 
higher $K$ (and masses), closer to our results (M.~Fedurco, in prep.).

The lines are clearly broadened by rotation, which itself is synchronized with
orbital period, despite the small but detectable eccentricity. 
Fast rotation of both components hampers our RV precision, thus the errors in masses,
which in this case are $\sim$5~per cent. The precision in absolute radii is only slightly
better (2-4~per cent), with error budget strongly influenced by uncertainties of 
fractional radii. The 4.3~mmag $rms$ of the LC is obviously caused by pulsations of 
the dSct type. In the Lomb-Scargle periodogram of the residuals
of \jkt fit we detect strong peaks at frequencies 15-21~d$^{-1}$, typical for
dSct stars. A separate publication with very detailed analysis of pulsations in 
KIC~4851217 is in preparation (by M.~Fedurco), therefore we did not tackle this 
problem in this work.

Most of the flux comes from the cooler, but more massive and larger secondary. The \isp
analysis of its spectrum (SNR$\sim$115) showed it to be an early F type star with
$T_{\rm eff,2}=7250(215)$~K. This is in agreement with e.g. $7306^{+225}_{-164}$~K from GDR2. 
We therefore expect the primary to be of spectral 
type A, with $T_{\rm eff}$ exceeding 8000~K. Unfortunately its disentangled spectrum
is not good enough for proper analysis (SNR$\sim$47). We made several attempts
to retrieve $T_{\rm eff,1}$ with \isp fixing as many parameters as possible 
(dynamical $\log(g)$, metallicity from secondary, etc.), and using model atmospheres 
appropriate for hotter stars, but we failed to reach such high temperatures.
From the secondary's spectrum we estimate the metallicity to be sub-solar, and 
without significant $\alpha$ enhancement. Without independent assessment of 
$T_{\rm eff,1}$ we can not estimate the distance using {\sc jktabsdim}.
We do this in Sect.~\ref{sec_iso} on the basis of isochrone-calibrated $T_{\rm eff,1}$.

\subsubsection{KIC~7821010}

\begin{figure*}
\includegraphics[width=0.9\columnwidth]{K0782_orb.eps}
\includegraphics[width=0.9\columnwidth]{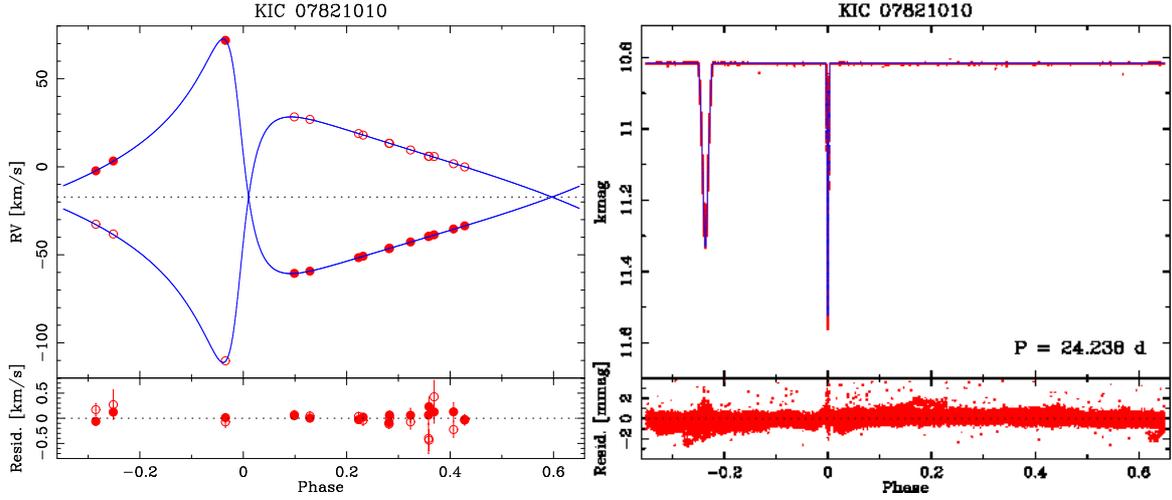}
\caption{Same as Fig. \ref{fig_mod_0343}, but for KIC~7821010.
}\label{fig_mod_0782}
\end{figure*}

This system has been reported on several occasions to harbour a circumbinary planet, 
detected with ETVs\footnote{e.g.: \tt http://www.astro.up.pt/investigacao/conferencias/\\ 
/toe2014/files/wwelsh.pdf} \citep{bor16}, but the proper publication is still to be announced 
(Fabrycky et al., in prep.). 
Since the publication of Paper II we obtained seven HIDES spectra of the system.
We have previously reported the uneven quality of the data, and that in favouring
observing conditions our observations result in SNR sufficient to obtain 
good precision RV measurements ($<$100 \ms). New observations aimed to increase the number
of ``good'' spectra, improve the precision of absolute stellar parameters, 
and possibly detect the RV signature of the planet, estimated from the solution of
\citet{bor16} to be 30-40 \ms, which is at the level of instrumental precision
in our programme (Paper~I). For the new orbital solution we did not use RVs coming 
from one of the older spectra, from May 05. 2015 (JD$\simeq$2457148.2), which had very 
low SNR ($<$15). Therefore, the final solution, and spectra disentangling, was based on 15
observations. The resulting TD spectra are of a good quality (SNR = 127 and 108 for 
the primary and secondary, respectively).

The short-cadence \kep photometry of this pair is available from Q02 and
Q09-Q17 (9 quarters). When comparing the light curves, we found that the eclipses in 
the long-cadence LC are slightly shallower and wider, likely because short-time-scale 
brightness variations were averaged out with the longer exposure time. 
For this reason our results base on the short-cadence data only.
We do not see significant out-of-eclipse variations, but we saw the influence 
of the circumbinary planet on $T_0$ and $P$ in each quarter. The fit
to a complete LC left characteristic residuals around eclipses, therefore we adopted the 
weighted averages from single-quarter fits. In Figure~\ref{fig_mod_0782} we present the
short-cadence LC, but the lower panel shows summarised residuals of single-quarter
fits. Some systematic residuals, coming from imperfect de-trending, are clearly
visible.

Mass uncertainties in the new solution (Tab.~\ref{tab_par_sb2}) are about 1/3 lower than in Paper~II 
($\sim$0.8~per cent for both components), and, thanks to improved $a\sin(i)$, also
the uncertainty in radii is quite good (0.8 and 1.1 per cent for the primary and 
secondary, respectively). The $rms$ of the orbital fit also decreased. With the selection of 
only the best RV measurements (10 for the primary and 8 for the secondary) we can 
lower it even further, to 55 and 60 \ms, respectively. This is, unfortunately, 
larger than the expected RV signal from the planet. While attempting to fit for this influence, 
basing on the orbital parameters from \citet{bor16}, we obtained a solution that was only 
marginally better (52+55 \ms\,$rms$). 

The \isp analysis resulted in similar effective temperatures of the components,
which is not surprising considering a mass ratio slightly lower than 1. The system's
metallicity is likely higher than solar (+0.10(8)~dex, with two individual results being
different by only 0.01~dex). We found no evidence for enhancement of $\alpha$-elements in neither 
component. With \isp we also fitted for the projected rotational velocities $v\sin(i)$. Lines of
both components are quite narrow, but some degree of broadening was expected for the secondary 
(from the width of cross-correlation function). Indeed, we found the secondary rotating faster 
(8.9$\pm$1.1~k\ms) than expected in pseudo-synchronous case (2.53$\pm$0.03~k\ms).
The formal error of the primary's $v\sin(i)$ is larger than the best-fitting value --
1.3$\pm$4.8~k\ms -- therefore the result is inconclusive. But from the secondary alone we
can conclude that KIC~7821010 did not reach the tidal equilibrium, therefore should be younger
than 3.6~Gyr, which is the time scale of synchronisation of rotation with orbital period 
(from {\sc jktabsdim}).

The two effective temperatures were used in \jktabs to estimate the distance $d_{\rm J}$.
The result -- 342(13)~pc -- is in a reasonable agreement with 360(5)~pc from GDR2. To ensure
the consistency between individual distances from various bands, we assumed $E(B-V)=0.11$~mag.
Without reddening, values corresponding to bands $B$ and $V$ are $\sim$60~pc larger than
those corresponding to $J,H,K$. The equivalent width of interstellar sodium D~1 line
is 0.21(1)~\AA, and correctly reproduces $E(B-V)\simeq0.11$~mag, according to calibrations by
\citet{mun97}.

\subsubsection{KIC~8552540 (V2277~Cyg)}
\begin{figure*}
\includegraphics[width=0.9\columnwidth]{K0855_orb.eps}
\includegraphics[width=0.9\columnwidth]{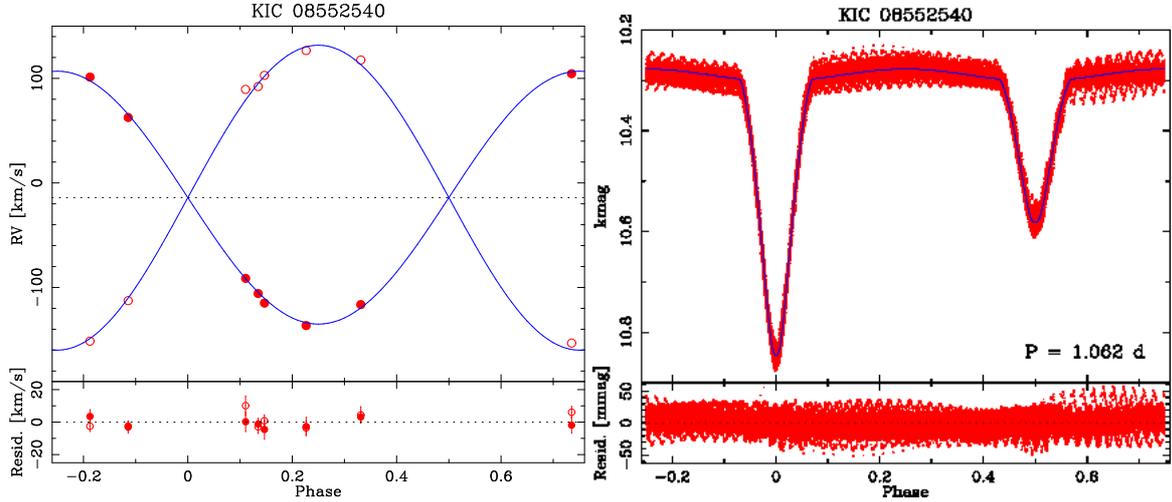}
\caption{Same as Fig. \ref{fig_mod_0343}, but for KIC~8552540.
}\label{fig_mod_0855}
\end{figure*}

\begin{table}
\centering
\caption{Comparison of our results for KIC~8552540 (V2277~Cyg) with parameters from 
\citet{mat17}.}\label{tab_k0855}
\begin{tabular}{lcc}
\hline \hline
	&This work$^a$	& \citeauthor{mat17}\\
Parameter & (Table \ref{tab_par_sb2}) & (\citeyear{mat17})\\
\hline
$P$ (d)		&    1.06193441(4)	& 1.06193426(-)	\\
$K_1$ (k\ms)	& 121.0(1.6)	& 121(1)	\\
$K_2$ (k\ms)	& 145.9(2.0)	& 153(2)	\\
$q$				&  0.829(16)	& 0.79(1)	\\
$i$ ($^\circ$)	&  85.83(46)	&  80.7(-)	\\
$a$ (R$_\odot$)	&  5.619(54)	& 5.83(5)	\\
$M_1$ (M$_\odot$)	& 1.153(36) & 1.32(3)	\\
$M_2$ (M$_\odot$)	& 0.956(28) & 1.04(2)	\\
$rms_1$ (k\ms)	& 3.0	& 4.1	\\
$rms_2$ (k\ms)	& 5.3	& 6.3	\\
\hline
\hline
\end{tabular}
\\$^a$ The same as in Paper~II.
\end{table}

The LC of this system is strongly affected by rapidly evolving spots. 
Short-cadence data are available only for fractions of quarters
Q02, Q03, and Q14, and are not enough to cover the evolution of spots properly and
obtain reliable results. We therefore use the fit to a complete Q00-Q17
long-cadence curve (shown in Fig.~\ref{fig_mod_0855}), with uncertainties calculated with 
our RS approach, i.e. the results from Paper~II remain intact. For the record, we give the 
$rms$ of the best fit to the short-cadence LC as well.

Furthermore, this system has no new HIDES observations, thus the only new addition is the 
\isp spectroscopic analysis. As KIC~4851217, this binary also has large 
ratio of component fluxes, and significantly different SNRs of disentangled 
spectra: 136 and 37 for the primary and secondary, respectively. It is the 
shortest-period binary in our sample, with two rapidly rotating components, 
therefore the final mass precision is relatively poor ($\sim$3 per cent).
The rapidly evolving spots affect the precision of the LC fit
(the largest $rms_{\rm LC}\simeq13$~mmag). It is worth noting that
the secondary may be considered a solar analogue. It has also been reported
in \citet{mat17}. The comparison is shown in Table~\ref{tab_k0855}.
There is an overall agreement, except $K_2$. This is not very surprising,
considering low precision of RV measurements of the faint and rotationally 
broadened secondary, which affected both the studies.

The \isp analysis of the primary's spectrum was done with $v\sin(i)$ fixed
to the value predicted by tidal locking. We ended up with [M/H$]=-0.27(11)$~dex
(lowest in our sample), no significant $\alpha$-element enhancement, and 
$T_{\rm eff,1}=6060(200)$~K. It agrees with $5870^{+290}_{-118}$~K from GDR2. 
Due to strong rotational broadening of lines, the uncertainties are relatively large. 
Spectrum of the secondary had its SNR too low for 
secure analysis, so without $T_{\rm eff,2}$ we are not able to calculate the 
distance with \jktabs at this stage. Attempts made on isochrone-based values
are described in Section~\ref{sec_iso}.

\subsubsection{KIC~9246715}
\begin{figure*}
\includegraphics[width=0.9\columnwidth]{K0924_orb.eps}
\includegraphics[width=0.9\columnwidth]{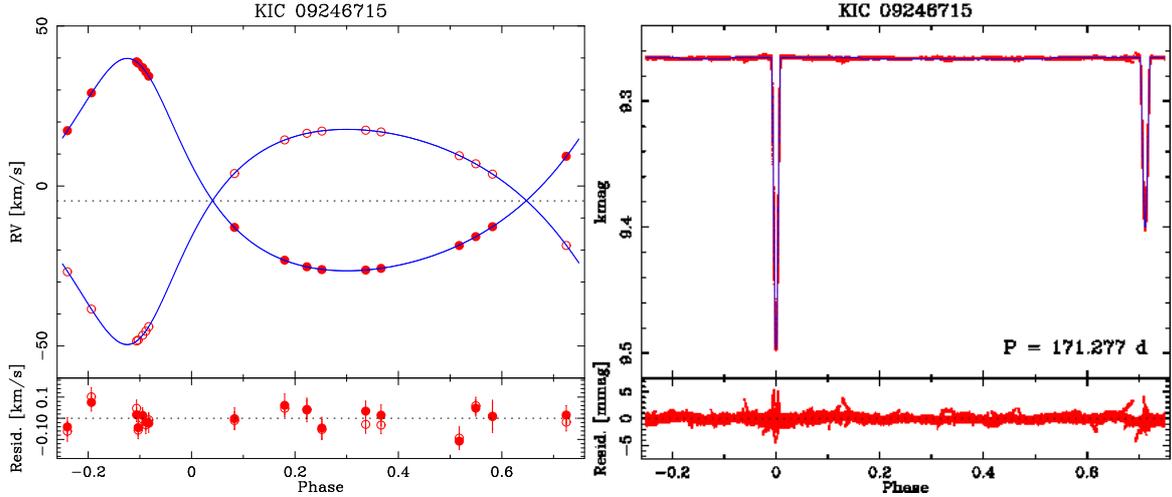}
\caption{Same as Fig. \ref{fig_mod_0343}, but for KIC~9246715.
}\label{fig_mod_0924}
\end{figure*}

\begin{table}
\centering
\caption{Comparison of updated results for KIC~9246715 with parameters from 
K924 and \citet{raw16}.}\label{tab_k924}
\begin{tabular}{lccc}
\hline \hline
	&This work	& K924 & \citeauthor{raw16}\\
Parameter & (Table \ref{tab_par_sb2}) & (Table 2) & (\citeyear{raw16})\\
\hline
$P$ (d)		& \multicolumn{2}{c}{171.2770(6)}	& 171.27688(1)	\\
$K_1$ (k\ms)	&  33.21(2)	&  33.18(16)	&  33.19$^{+0.04}_{-0.05}$	\\
$K_2$ (k\ms)	&  33.63(2)	&  33.58(14)	&  33.53$^{+0.04}_{-0.05}$	\\
$e$				& 0.3552(4)	&  0.3587(9)	&  0.3559$^{+0.0002}_{-0.0003}$	\\
$r_1$		& \multicolumn{2}{c}{0.04008(58)}	& 0.0396$^{+0.0001}_{-0.0003}$	\\
$r_2$		& \multicolumn{2}{c}{0.03870(40)}	& 0.0393(1)	\\
$i$ ($^\circ$)	&\multicolumn{2}{c}{87.049(31)}	& 87.051$^{+0.009}_{-0.003}$	\\
$M_1$ (M$_\odot$)	& 2.187(3)   & 2.169(24)	& 2.171$^{+0.006}_{-0.008}$	\\
$M_2$ (M$_\odot$)	& 2.160(3)   & 2.143(25)	& 2.149$^{+0.006}_{-0.008}$	\\
$R_1$ (R$_\odot$)	& 8.49(12)	 &  8.47(13)	& 8.37$^{+0.03}_{-0.07}$	\\
$R_2$ (R$_\odot$)	& 8.20(9)    &  8.18(9) 	& 8.30$^{+0.04}_{-0.03}$	\\
$T_{\rm eff,1}$ (K) &  4890(50)  & ---		 	& 4990(90)	\\
$T_{\rm eff,2}$ (K) &  4905(55)	 & ---		 	& 5030(80)	\\
$[$M/H$]_1$			&	-0.01(2) & ---		 	& -0.22(12)$^a$	\\
$[$M/H$]_2$			&	 0.03(2) & ---		 	& -0.10(9)$^a$	\\
$rms_{\rm RV1}$ (k\ms) & 0.046 & 0.045 & 0.55$^b$ \\ 
$rms_{\rm RV2}$ (k\ms) & 0.051 & 0.052 & 0.55$^b$ \\ 
\hline
\end{tabular}
\\$^a$ [Fe/H] obtained with {\sc moog}.
\\$^b$ Not given directly in \citet{raw16}, therefore it was calculated in this work with \vfit using their exact solution.
\end{table}

This system was previously described in a dedicated paper (K924), and since then
it has been observed 9 more times with HIDES, making the total number of spectra 17.
No \kep short-cadence data are available. Except eclipses the long-cadence curve  
shows systematics, presumably coming from imperfect de-trending, also affecting some of
the eclipses. It should be noted that the long period of this system (171~d) causes that 
both minima are not always recorded during one quarter. The approach to the LC fit was, 
therefore slightly different, and is described in details in K924. We do not repeat the
LC analysis here.

KIC~9246715 is one of the most interesting cases in our sample. It is composed of two red giants,
one of which shows solar-type oscillations. At the time of publication it was only the 
third example of a Galactic double-giant eclipsing binary with masses and radii 
measured with precision below 2~per cent. It was simultaneously and independently analysed 
by \citet{raw16}, who used their own set of 24 high-resolution spectra (from TRES, ARCES,
and APOGEE), and derived parameters in agreement with K924. Comparison of these results
with K924 and this work is shown in Table~\ref{tab_k924}. The full set of our parameters
is presented in Table~\ref{tab_par_sb2}. Additionally, \citet{raw16} estimated 
temperatures and iron abundance from the disentangled component spectra, analysed pulsations,
and compared the binary with models. They concluded that one of the components is in a
helium (He) burning phase.

The fact that KIC~9246715 contains a solar-type oscillator makes it useful for testing 
and calibrating the asteroseismic relations \citep{gau16,bro18}. The analysis of oscillations
is dependent on the evolutionary status of a star, which requires very precise observables 
to be securely established. One should note that in K924 there were only 8 RV measurements 
used, and resulting precision of $K_{\rm 1,2}$ severely suffered from that fact, even though
the $rms$ of the orbital fit was only $\sim$50~\ms (close to the stability level of HIDES
in our survey; Paper~I). Also, the spread of RV residuals in \citet[][Fig.~6]{raw16} reaches
$\pm$2~k\ms, suggesting the $rms$ of the order of few hundred \ms, and clearly shows
systematic effects (residuals for one component strongly correlated with the other). 
It is therefore likely that their uncertainties of $K_{\rm 1,2}$ (40-50 \ms) are
underestimated, as are the errors of masses.
We followed KIC~9246715 in order to improve the orbital, and therefore physical parameters 
of its components. 

We doubled the number of HIDES spectra, but the quality of
the orbital fit remained almost unchanged. With more data and better sampling we
downed the systematics in velocity amplitudes $K_{1,2}$, and
also improved our estimate of eccentricity, which, as it turns out, has large influence 
on the final errors of masses. The final precision in $\sim$0.15~per cent for both 
components. The precision in $R$ is 1.4 + 1.0 per cent, analogously.
Results for radii did not change much, as we did not re-analyse the LC.

Both disentangled spectra have decent SNR (123 and 117), and their \isp analysis gave
similar temperatures of components: 4890(50) and 4905(60)~K for the primary and secondary,
respectively. Individual values of [M/H] and [$\alpha$/Fe] were in good agreement, and
pointed towards chemical composition indifferent from solar. Rotational velocities $v\sin(i)$ 
were fitted for, and we found them to be in very good agreement with those 
predicted by pseudo-synchronisation: 2.51(4) and 2.42(3)~k\ms (from {\sc jktabsdim}).
It is, however, unlikely that at such separation tidal forces influenced the rotation
and synchronised it with the orbital period during the life time of the system.
The observed rotation is rather primordial. 

We used \jktabs and our values of $T_{\rm eff}$s to estimate the distance. We found it to be
573(13)~pc, assuming $E(B-V)\simeq0.13$~mag, which is significantly less than 616(11)~pc 
from GDR2. When no reddening is assumed, individual distances in $B$ and $V$ are larger
from those in $J,H,K$ by $\sim$150 and 100~pc, respectively. The IR values themselves are all
around 600~pc, still too low for GDR2.

\subsubsection{KIC~9641031 (FL Lyr)}
\begin{figure*}
\includegraphics[width=0.9\columnwidth]{K0964_orb.eps}
\includegraphics[width=0.9\columnwidth]{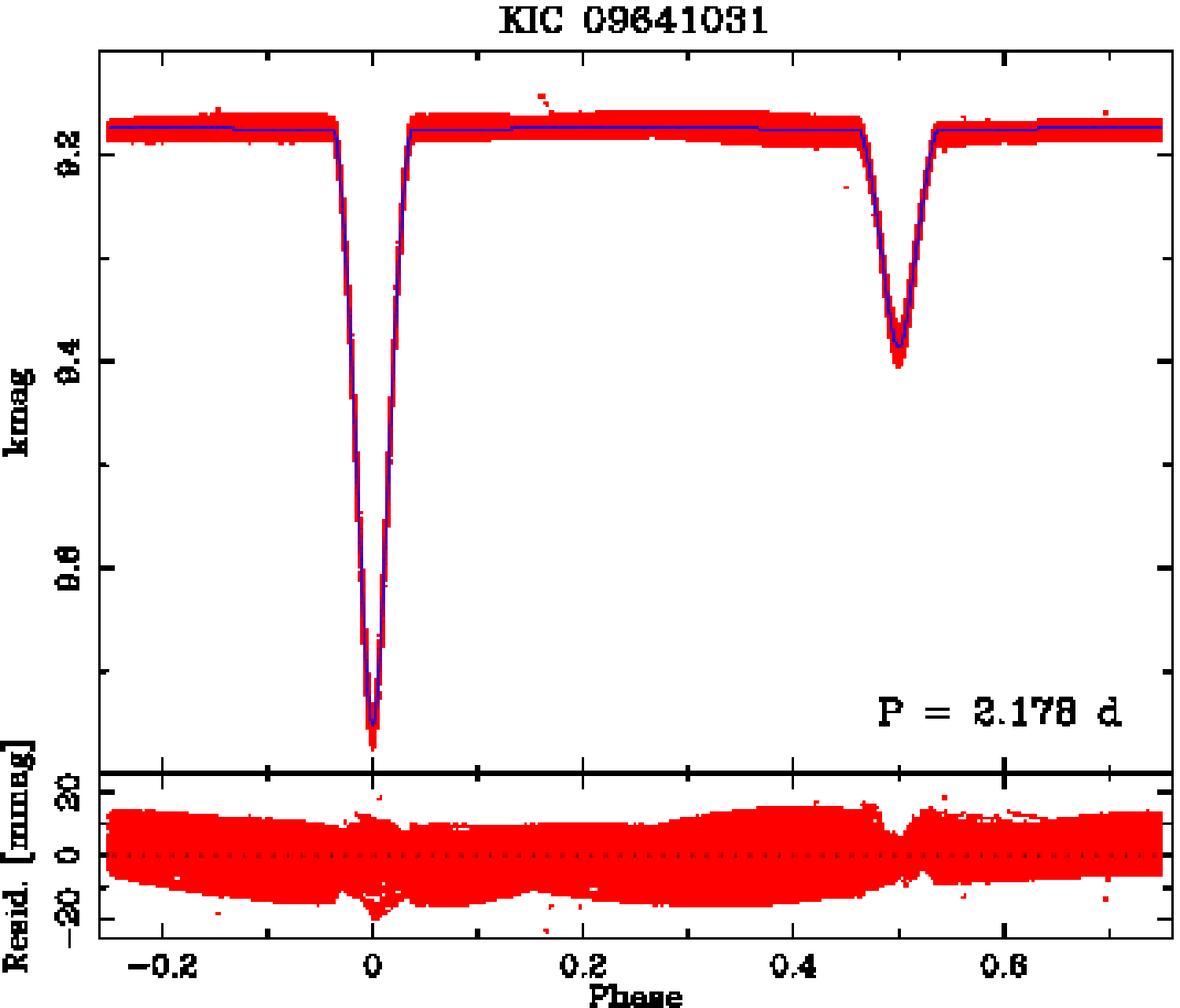}
\caption{Same as Fig. \ref{fig_mod_0343}, but for KIC~9641031.
}\label{fig_mod_0964}
\end{figure*}

\begin{table}
\centering
\caption{Comparison of updated results for KIC~9641031 (FL Lyr) with parameters from 
Paper~II and \citet{pop86}.}\label{tab_fllyr}
\begin{tabular}{lccc}
\hline \hline
	& This work	& Paper~II & \citeauthor{pop86}\\
Parameter & (Table \ref{tab_par_sb2}) & (Table 2) & (\citeyear{pop86})\\
\hline
$P$ (d)		& \multicolumn{2}{c}{2.17815425(7)}	& 2.1781542(3)	\\
$K_1$ (k\ms)	&  93.34(10)	 &  93.23(12)	&  93.5(5)	\\
$K_2$ (k\ms)	& 118.75(35)	 & 118.19(30)	& 118.9(7)	\\
$r_1$			&  0.1361(25)	 & 0.1389(25)	& 0.140(3)	\\
$r_2$			&  0.0984(26)	 & 0.0995(27)	& 0.105(3)	\\
$i$ ($^\circ$)	&  87.13(71)	 & 85.36(71)	& 86.3(4)	\\
$M_1$ (M$_\odot$)	& 1.2102(76) & 1.2041(76)	& 1.218(16)	\\
$M_2$ (M$_\odot$)	& 0.9512(39) & 0.9498(46)	& 0.958(11)	\\
$R_1$ (R$_\odot$)	& 1.244(23)  &  1.269(23)	& 1.283(30)	\\
$R_2$ (R$_\odot$)	& 0.900(24)  &  0.908(24)	& 0.963(30)	\\
$T_{\rm eff,1}$ (K) &  6264(112) & ---		 	& 6150({\it 100})	\\
$T_{\rm eff,2}$ (K) &  5490(247) & ---		 	& 5300({\it 100})	\\
$[$M/H$]$ 			&	-0.07(9) & ---		 	& 0.32(--)	\\
\hline
\end{tabular}
\end{table}

This binary has a very extensive set of short-cadence data points,
starting in quarters Q01 and Q02, through almost entire quarters Q07 and Q08,
and with a nearly continuous coverage since Q13 till the end of the mission.
In this case, we also noted slightly different eclipse depths in short- and
long-cadence data, therefore the final values come from a fit to a complete
short-cadence curve (also in Figure~\ref{fig_mod_0964}), although the long-cadence
curve gives similar results. However, both components of this pair seem to have 
prominent, rapidly evolving spots, which make the shape of the LC changing in 
relatively short time scales (a significant change is seen after only several orbits). 
The complete long-cadence light curve, whose coverage in time is much more 
complete\footnote{There are no \kep data from Q04.} seems to probe the evolution 
of spots more uniformly, and their influence on the phase-folded LC appears to
be averaged out. This kind of situation favours the RS method over the MC for
error determination, therefore, the uncertainties quoted in Table~\ref{tab_par_sb2}
come from the combined RS+$rms$ analysis of single-quarter long-cadence data.

We observed this system with HIDES three more times since Paper~II, 
increasing the number of spectra to 15 (14 of which were used in TD). 
The new data were taken mainly to test the hypothesis of a low-mass circumbinary 
body on a 103-day orbit, based on ETVs from Paper~II. We do not detect any RV modulation 
at the expected level ($\sim$1.26 k\ms), therefore we conclude that the observed
ETVs were likely caused by evolution of spots. The overall precision of RV data
and level of uncertainties are similar to those from Paper~II (0.5-0.7~per cent 
in mass, 1.8-2.6 per cent in radii).
In principle, in comparison to the previous extensive study of this system by
\citet{pop86}, we reach 2-3 times lower errors in mass, and comparable 
(though slightly lower) in radii. To decrease errors of $R$ one would have to
remove the influence of spots on the LC, but for this, a knowledge about their
location (which component is affected) is required, in order not to hamper the
depths of the eclipses. Rapid evolution of spots on this system makes such
an analysis quite challenging. 

Our updated parameters are in better agreement with \citet{pop86} than in 
Paper~II, due to larger $K_2$. Direct comparison is presented in 
Table~\ref{tab_fllyr}.
We now also add effective temperatures and metallicity of KIC~9641031, 
obtained from spectra. \citeauthor{pop86} estimated $T_{\rm eff}$s from the 
$V-R$ colours from their photometric solution and on the basis of calibrations 
from \citet{pop80}, and metal content $Z=0.04$ from comparison with isochrones on 
mass-temperature plane. They also admit that not all uncertainties 
were included in temperature errors. Our $T_{\rm eff}$ estimates from \isp
are in a formal agreement (due to error bars) but both are larger.
When compared to a modern calibration, from \citet{wor11}, the adopted
$V-R$ indices (0.46 and 0.61 mag for the primary and secondary, respectively)
actually predict temperatures about 1000~K lower, or themselves are 0.15 mag 
too high. One should also note that the calibrations from \citet{pop80} played
a role in establishing the ratios of radii $k$ and fluxes in \citet{pop86}.
Our approach is calibration-independent, and resulted in different
values of $k$ and (subsequently) $R_2$. The issue with uncertainties 
of $r$ and $R$ in \citet{pop86} was already raised in Paper~II.
For these reasons we advise to treat the results from \citet{pop86}
with caution. The difference in metallicity will be discussed in Sect.~\ref{sec_iso}.

The temperatures from Tab.~\ref{tab_par_sb2} were used to estimate the distance
with \jktabs. The resulting value of 130(5)~pc is in a good agreement with 135.0(5)~pc
from GDR2, but a small amount of reddening ($E(B-V)=0.04$~mag) had to be assumed to reach
consistency between distances from $B$ and $V$ bands with those from $J,H,K$.
Without reddening, the $B,V$ distances were systematically larger by 5-10~pc.

\subsubsection{KIC~10031808}

\begin{figure*}
\includegraphics[width=0.9\columnwidth]{K1003_orb.eps}
\includegraphics[width=0.9\columnwidth]{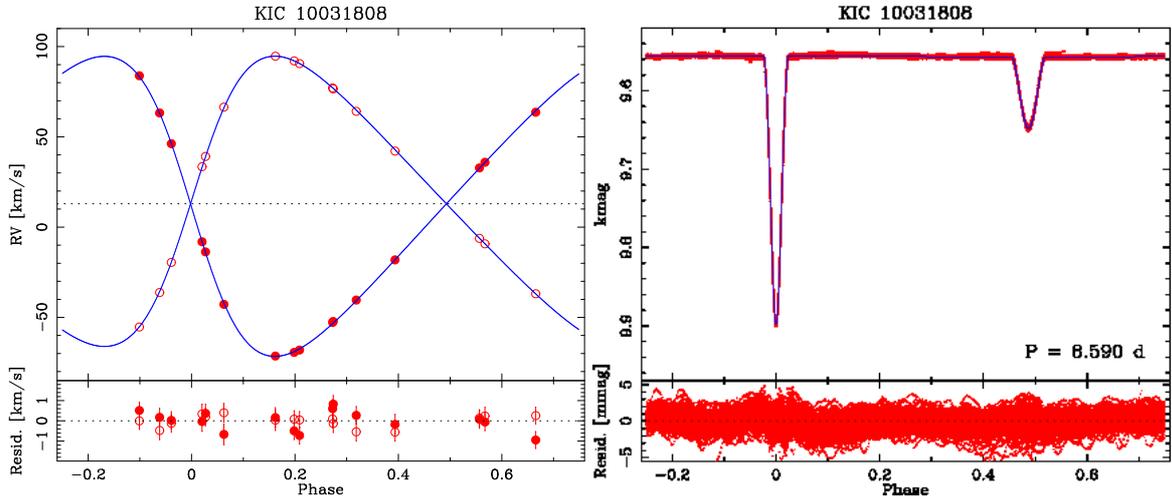}
\caption{Same as Fig. \ref{fig_mod_0343}, but for KIC~10031808.
}\label{fig_mod_1003}
\end{figure*}

In the LC analysis of KIC~10031808 we did not use short-cadence data,
as they are only available for a fraction of quarter Q02. In general, the shapes and
depths of eclipses agree in both short- and long-cadence photometry. Figure~\ref{fig_mod_1003}
depicts the long-cadence curve, and the final parameters were taken from a fit to a complete
Q00-Q17 set, i.e. results presented in Paper~II remain unchanged.

This system has no new HIDES observations since Paper~II, but is one of the 
most interesting targets in the sample. It contains a $\gamma$~Doradus ($\gamma$Dor) type pulsator, 
and is one of only few known cases of such a star in an eclipsing binary with precisely 
measured parameters (i.e. 0.4-0.8 per cent in both masses and radii). The DEBCat lists
only two systems with $\gamma$Dor stars: CoRoT~102918586 \citep{mac13} and KIC~11285625 
\citep{deb13}. There are, of course, other cases known, but they do not have
parameters derived with sufficient precision. Comparison of
our values of $T_{\rm eff}$, $\log(g)$, and $v_{\rm mic}$ of both components with the 
distributions of those parameters in \citet{kah16} suggests the 1.74~M$_\odot$ primary 
is the pulsator. In particular, in \citet{kah16} there is no case of a $\gamma$Dor pulsator 
with $\log(g)$ lower than 3.8~dex.

We found the metallicity of this system to be slightly sub-solar ($-0.11\pm0.08$~dex), 
but both components shown a clear $\alpha$-enhancement ($0.16\pm0.06$~dex). 
The obtained effective temperatures are close to those predicted in Paper~II, 
but the resulting $d_{\rm J}=445(15)$~pc is only in a $\sim$2$\sigma$ agreement with 
one from GDR2. Our distance estimate assumes $E(B-V)=0.125$~mag, which is required to 
obtain consistency between distances calculated for each band. A better match with GDR2 
is found when $E(B-V)\simeq0.08$~mag, but in such case distances from $B$ and $V$ 
bands are systematically larger than those from $J,H,K$ by about 30~pc. The equivalent 
width of sodium D1 line is 0.43(5)~\AA, and, according to calibrations from \citet{mun97}, 
favours the value of $E(B-V)=0.125$~mag.

\subsubsection{KIC~10191056~A}\label{sec_1019A}

\begin{figure*}
\includegraphics[width=0.9\columnwidth]{K1019_orb.eps}
\includegraphics[width=0.9\columnwidth]{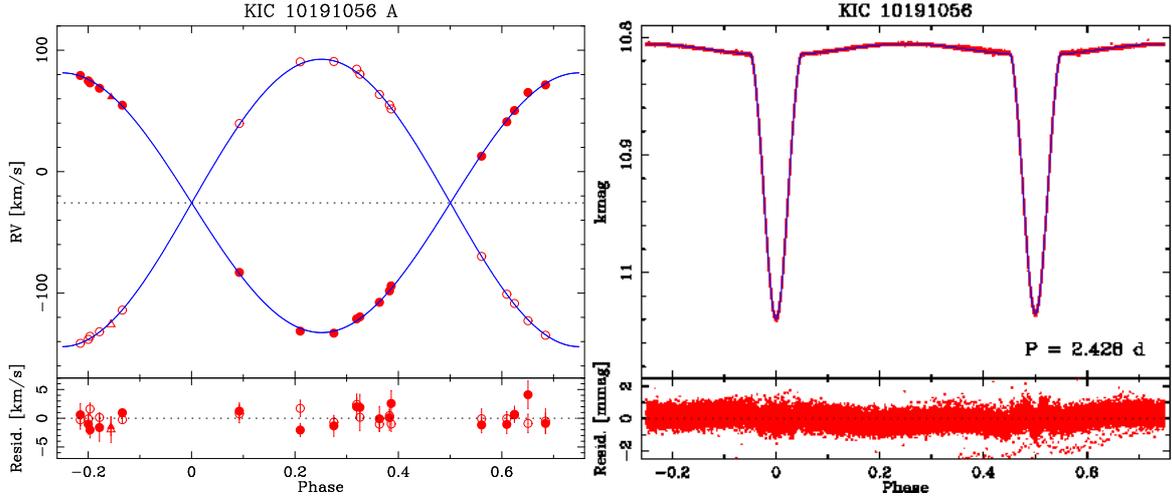}
\caption{Same as Fig. \ref{fig_mod_0343}, but for KIC~10191056~A and
RV measurements from TRES (triangles).
}\label{fig_mod_1019}
\end{figure*}

\begin{table}
\centering
\caption{Comparison of updated results for KIC~10191056 with parameters from 
Paper~II and \citet{mat17}.}\label{tab_k1019}
\begin{tabular}{lccc}
\hline \hline
	&This work	& Paper~II & \citeauthor{mat17}\\
Parameter & (Table \ref{tab_par_sb2}) & (Table 2) & (\citeyear{mat17})\\
\hline
$P$ (d)			& \multicolumn{2}{c}{2.42749488(2)}	& 2.42749484(-)	\\
$K_1$ (k\ms)	& 106.96(58)	& 107.0(1.3)	& 100(2)	\\
$K_2$ (k\ms)	& 118.34(33)	& 119.3(1.0)	& 119(2)	\\
$q$				&  0.9038(55)	&  0.897(13)	& 0.83(2)	\\
$i$ ($^\circ$)	&  81.33(8) 	&  81.35(8) 	& 80.5(-)	\\
$a$ (R$_\odot$)	&  10.938(33)	&  10.986(80)	& 10.7(2)	\\
$M_1$ (M$_\odot$)& 1.564(12)	&  1.590(32)	& 1.50(5)	\\
$M_2$ (M$_\odot$)& 1.413(16)	&  1.427(36)	& 1.25(4)	\\
$rms_{\rm RV1}$ (k\ms) & 1.7	&  2.5	& 6.8 \\ 
$rms_{\rm RV2}$ (k\ms) & 1.1	&  2.3	& 6.9 \\ 
\hline
\end{tabular}
\end{table}

The triple-lined KIC~10191056 was observed 8 additional times since Paper~II,
but here we did not use low-SNR spectrum taken on October 10, 2016
(JD$\simeq$2547671.97), thus the total number of HIDES spectra is 18.
The main goal was to monitor the variations of systemic velocity $\gamma$ of
the eclipsing pair A, and the motion of the tertiary B, which will be 
described in details in Section~\ref{sec_k1019b}. The TRES spectrum was included
in this study because of the time between it was taken and our first 
observations (nearly 450 days), so we hoped to detect long-term variations
with higher significance. Analysis of the newly added data confirmed
that there is no significant variation of $\gamma$ in time, which was
already proposed in Paper~II. When fitted for, the linear RV trends for both
components were $\sim$10 times smaller than their uncertainties. We can put a
formal upper limit on the linear trend in $\gamma$ at 0.8~\ms~d$^{-1}$.
Notably, the RVs from TRES spectrum matched the HIDES 
data very well even without applying any zero-point shift (Fig.~\ref{fig_mod_1019}).
The total time-span is now 1182~d with TRES, and 737~d without.
The meaning of the RV variations of component B will be discussed later.

In Table~\ref{tab_par_sb2} we present updated orbital and physical parameters
of KIC~10191056~A, under assumption of constant $\gamma$, and utilising new
\jkt results. The long-cadence LC shows shallower eclipses, the amount and time
span of the short-cadence data is sufficient (quarters Q02, Q04 and Q06-Q10),
and no obvious spot-like modulation or pulsations are present,
thus the usage of short-cadence data is preferable and justified. We adopted
results of the fit to the complete short-cadence set, and this is the LC shown
in Figure~\ref{fig_mod_1019}.

With respect to Paper~II we have significantly reduced
the errors in $K_{1,2}$ and dependent parameters, i.e. the relative mass uncertainties
are now 0.8 and 1.1 per cent for the primary and secondary, respectively. The 
errors in radii are now slightly better as well -- 1.2 and 1.6 per cent.
In Table~\ref{tab_k1019} we compare our current and previous results with those of 
\citet{mat17}. We note a significant discrepancy in $K_1$, which leads to 
disagreement in other parameters. It could be
caused by the fact that \citet{mat17} used few observations in phases around eclipses,
when the RV difference is small, and lines are blended.

Without the spectral analysis, we do not have independent 
$T_{\rm eff}$ and [M/H] estimates.

\subsubsection{KIC~10583181}
\begin{figure*}
\includegraphics[width=0.9\columnwidth]{K1058_orb.eps}
\includegraphics[width=0.9\columnwidth]{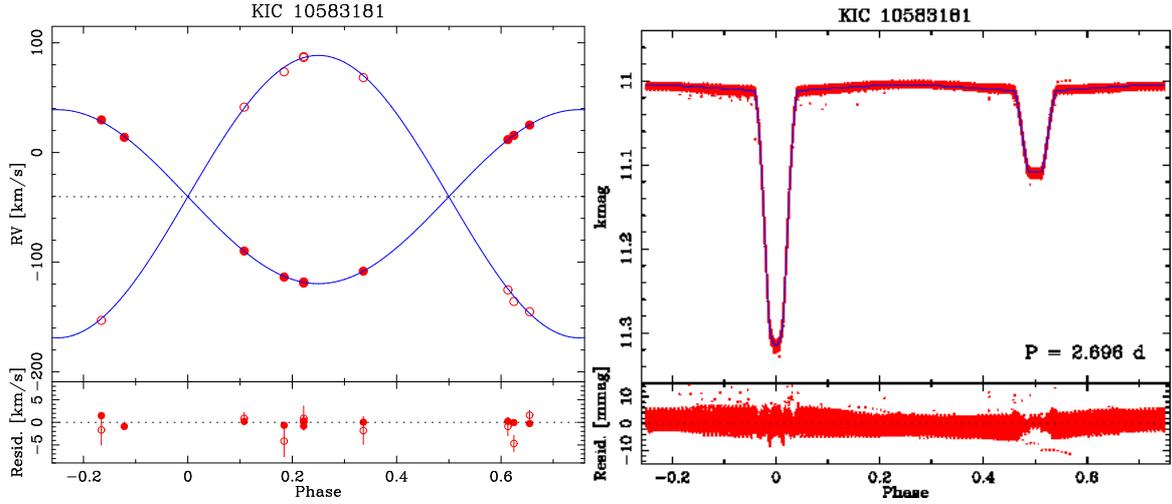}
\caption{Same as Fig. \ref{fig_mod_0343}, but for KIC~10583181. The RV
modulation coming from the circumbinary body has been removed.
}\label{fig_mod_1058}
\end{figure*}

\begin{figure}
\includegraphics[width=0.9\columnwidth]{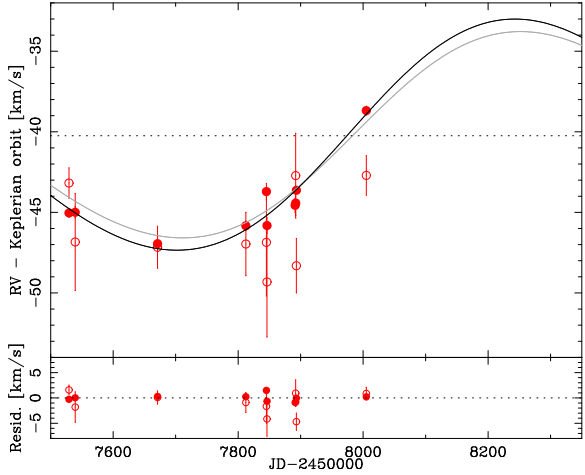}
\caption{RVs of KIC~10583181 as a function of time after removing the inner Keplerian orbit. 
Symbols are the same as in Fig.~\ref{fig_mod_1058}. The black line is 
the best-fitting model of centre of mass velocity $\gamma$ variation.
The grey line is $\gamma$ variation predicted by the exact solution from
\citet{bor16} The plot is stretched to JD=2458350 to show the total scale
of velocity variation.
}\label{fig_1058_cb}
\end{figure}

This system was observed 10 times with HIDES, but we acquired only 9 measurements
of the faint secondary, therefore 9 spectra were used in TD. In the orbital fit 
with \vfit we took into account the presence of the circumbinary body reported by
\citet{bor16}. The circumbinary orbit was assumed to be Keplerian, and parametrized 
by period $P_3$, eccentricity $e_3$, moment of pericentre passage $T_3$, longitude of
pericentre $\omega_3$, and semi-amplitude of modulation of the inner binary's systemic
velocity $K_3$. In such mode, the parameter $\gamma$ in \vfit is defined as the 
systemic velocity of the whole triple system. Parameters of the outer orbit were
found simultaneously with those of the inner binary.

The time span of our HIDES observations is 482~d, while $P_3=1169.2$~d, therefore we
could not set the outer period free in our fit (although we have data taken around the
pericentre passage). We also fixed $e_3=0.06$, and $\omega_3=279^\circ$ -- values given by 
\citet{bor16}\footnote{In \vfit $\omega_3$ is defined for the outer body, while in \citet{bor16}
for the inner binary, therefore we added $\pi$ (180$^\circ$) in our fit.}, and only searched for $K_3$ and $T_3$.
Resulting parameters of the inner orbit (only) are given in Table~\ref{tab_par_sb2}. 
Figure~\ref{fig_mod_1058} shows only the RV
curves of the inner binary, while the modulation of $\gamma$ induced by the third
body is shown in Fig.~\ref{fig_1058_cb}. 

\citet{bor16} do not give the value of amplitude of the ETVs, but list the $a\sin(i_3)$, from
which we can estimate the expected scale of RV variation. From $a\sin(i_3)=154.0(1)$~R$_\odot$,
and using $P_3$ and $e_3$ given above, we expect $K_3$ to be 6.674(7)~k\ms. Our 
orbital solution gives 7.2(1.6)~k\ms, in agreement with the expected one. For the record,
our value of $T_3 = 4494(53)$~d (JD-2450000) is also in agreement with the value from 
\citet{bor16} -- 4503(6)~d. Larger errors are, of course, the effect of poor time
coverage of the outer orbit. The minimum companion mass, from our solution, is
0.66(14)~M$_\odot$.

Overall, our orbital solution is satisfactory, considering relatively poor 
precision of the RVs. Our measurements are hampered by fast rotation (36 and 19 k\ms
for the primary and secondary, respectively), and small contribution of the secondary
to the total flux ($\sim$10~per cent). Yet, the $rms$ of the fit for the primary is 
only 0.7~k\ms. Relative mass uncertainties are also quite low: 2.3 and 1.7~per cent 
for primary and secondary, respectively.

In Figure~\ref{fig_mod_1058} we also show the short-cadence \kep light curve, taken
during quarters Q02, Q03, and Q7-Q10.
Due to the presence of the third body, which strongly influences moments of eclipses,
for the LC fit we used a similar approach as for KIC~7821010, and analysed the
data quarter-by-quarter. Additionally, we noted variations of the depth of minima,
especially the flat (total) secondary. The lower panel of the LC fit in Fig.~\ref{fig_mod_1058}
shows residuals from single-quarter fits stacked together. A prominent feature is that
the scatter during the secondary minimum is much lower than outside of eclipses.
This suggests that the secondary, which turns out to be a solar-analogue, contains
cool spots which evolve in time. There are also several flares recorded in the LC.

We also fitted for the contribution of the third light $l_3/l_{\rm tot}$ and we found 
it to be 0.12(2), i.e. larger than $l_2$ ($l_2/l_{\rm tot}\simeq0.084$), although
probably variable ($>$0.12 in Qs: 02, 07, 09, and 10; $<$0.11 in Qs 03 and 08). 
The variability may be a result of the satellite's positioning and rotation, as 
for KIC~3439031, but the total amount of $l_3$ is difficult to explain by contamination 
from nearby stars, so the bulk of it must come from the circumbinary companion.
We do not, however, see this companion in our spectra. Considering a relatively
large minimum mass, this can be explained if the companion was a binary itself,
composed of two M or K-type stars, too faint to be detected with our approach.
However, with our current data we can not verify this.

The overall LC fit is quite good ($rms\simeq1.7$~mmag), hampered mildly by the
evolution of spots, and led to relatively low uncertainties of radii: 0.7 and 1.0
per cent for the primary and secondary, respectively. We note that the primary
is very similar, only slightly smaller, than the primary of KIC~10191056. The orbital
periods and major semi-axes are also alike, both pairs accompanied by other bodies,
and the main difference between them are, obviously, the secondaries. Moreover,
the secondary in KIC~10583181 has almost identical mass as the secondary in KIC~8552540. 
It would be interesting to compare a number of such pairs (identical primaries with different 
secondaries, or vice versa) and search for differences that could be caused by influence of
different companions, e.g. interaction of magnetic fields, influence of rotation on
the internal structure and activity, etc.

The \isp analysis was possible only for the primary. The influence of the third light
was assumed to be constant across the spectrum, which is probably not entirely
correct, but, given that we have no additional information about $l_3$, was the
only reasonable assumption we could make. Depths of spectral lines were probably affected
in a non-uniform way, therefore the results of this analysis should be treated with caution.
We found that the system's chemical composition is probably similar to solar, and the primary 
is cooler than expected for the Main Sequence (MS). Our $T_{\rm eff,1}=6730(140)$~K is, however,
larger than $6425^{+260}_{-180}$~K from GDR2, although formally in 1$\sigma$ agreement . 
We found no literature estimates of $T_{\rm eff}$ that would match values expected at MS 
for this mass and solar metallicity. At the same time, the primary's radius $R_1$ is 
larger than predicted for the zero-age MS, which suggests it has evolved significantly. 
More detailed discussion is presented in Sect.~\ref{sec_iso}.

Without both effective temperatures, and information about the third light in bands
other than {\it Kepler}'s, we are unable to estimate $d_{\rm J}$. We estimate the 
distance on the basis of isochrone-calibrated absolute magnitudes $M_{\rm kep}$ in Sect.~\ref{sec_iso}.

Finally, we compare our direct determination of absolute masses with
indirect approach from \citet{dev08}. By comparing parameters obtained 
from analysing the TrES light curve (T-Lyr-01013) and apparent magnitudes 
of the whole system with theoretical isochrones, \citeauthor{dev08} obtained 
the most probable masses (and age) of the components: 1.749(19) and 
1.049(15)~M$_\odot$. The disagreement with absolute values from 
Tab.~\ref{tab_par_sb2} is obvious, but, surprisingly, the mass ratio agrees. 
This is another case (more are given in Paper~II) which shows that without 
any spectroscopic information, it is impossible to obtain reliable stellar 
parameters for eclipsing binaries.

\subsubsection{KIC~10987439}
\begin{figure*}
\includegraphics[width=0.9\columnwidth]{K1098_orb.eps}
\includegraphics[width=0.9\columnwidth]{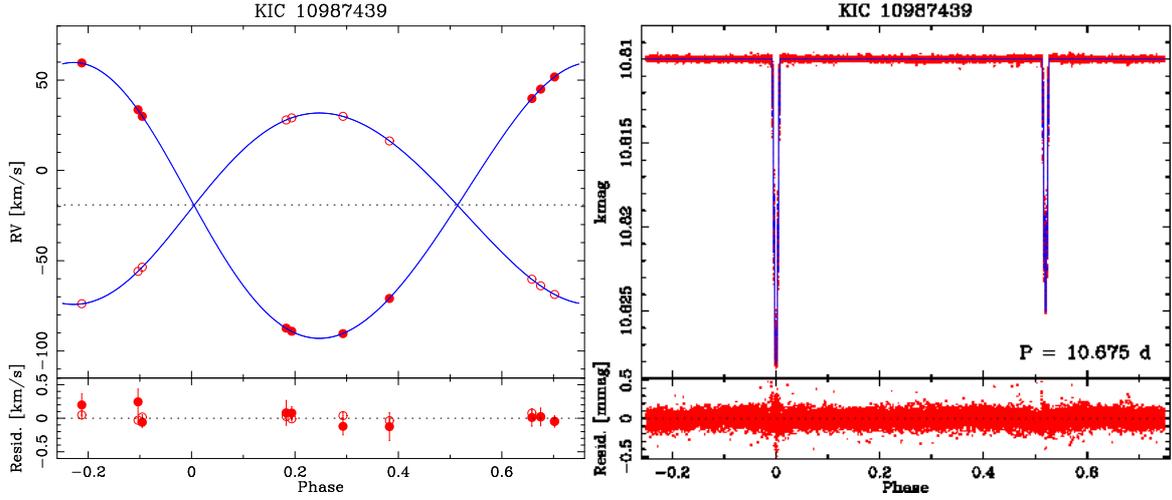}
\caption{Same as Fig. \ref{fig_mod_0343}, but for KIC~10987439.
}\label{fig_mod_1098}
\end{figure*}

\begin{figure}
\includegraphics[width=0.9\columnwidth]{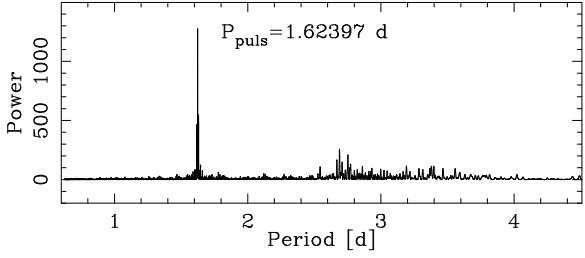}
\caption{Lomb-Scargle periodogram of residuals of our \jkt fit
to the \kep long-cadence curve of KIC~10987439. The detected periodicities,
with the dominant one at 1.624 days, suggest $\gamma$Dor-type pulsations.}\label{fig_per_1098}
\end{figure}

This system has no new HIDES observations since Paper~II, and only long-cadence
\kep photometry is available, thus the we rely on our previous results.
It is another case of a high-contrast
pair (low flux ratio), for which the disentangled spectrum of the fainter component is not 
sufficient for reliable spectroscopic analysis (SNR=20). It is, however, an important target, 
since we reached excellent precision in masses (0.34 and 0.32 per cent), and also
very good in radii (1.6 and 2.0 per cent for the primary and secondary, respectively).
Please note that, as in KIC~4851217 or 10031808, the primary is actually the fainter star.

The \isp analysis of the TD spectrum of the secondary (SNR=134) gave the 
$T_{\rm eff,2}=6490(90)$, the systems metallicity and $\alpha$-enhancement 
indistinguishable from solar: $-0.03(5)$ and 0.05(16)~dex, respectively. 
The temperature is in excellent agreement with $6450^{+270}_{-25}$~K from GDR2. 
During the analysis we assumed
pseudo-synchronous rotation, because the predicted time scale of synchronisation (from 
{\sc jktabsdim}) is only 93~Myr. Again, we could not use \jktabs to calculate distance.
We do it on the basis of isochrone-calibrated values in Sect.~\ref{sec_iso}.

However, in addition to Paper~II, in this work we run a Lomb-Scargle periodogram
on the residuals of our fit. We detected a single, dominant peak at
$\simeq1.624$~d with two sidelobes at $\Delta P\simeq \pm0.0068$~d 
(or $\Delta \nu \simeq \pm 0.00258$~d$^{-1}$ in frequency domain), and a variety
of lower, but still very prominent peaks at periods between 2.5 and 4 days
(Figure~\ref{fig_per_1098}). The amplitude of the dominant pulsation is only 33~ppm.
We tentatively interpret this as $\gamma$Dor-type pulsations, originating most likely from 
the brighter, more massive secondary, but a detailed analysis is required. 
Comparison of our values of $T_{\rm eff}, \log(g)$, and $v_{\rm mic}$ with distributions
of those parameters in \citet{kah16} tends to confirm our conclusion, although 
KIC~10987439 resides at the lower-mass edge of distributions. If pulsations are 
confirmed, KIC~10987439 together with KIC~10031808, would double the number of precisely 
measured eclipsing binaries with $\gamma$Dor-type stars. 

\subsubsection{KIC~11922782}
\begin{figure*}
\includegraphics[width=0.9\columnwidth]{K1192_orb.eps}
\includegraphics[width=0.9\columnwidth]{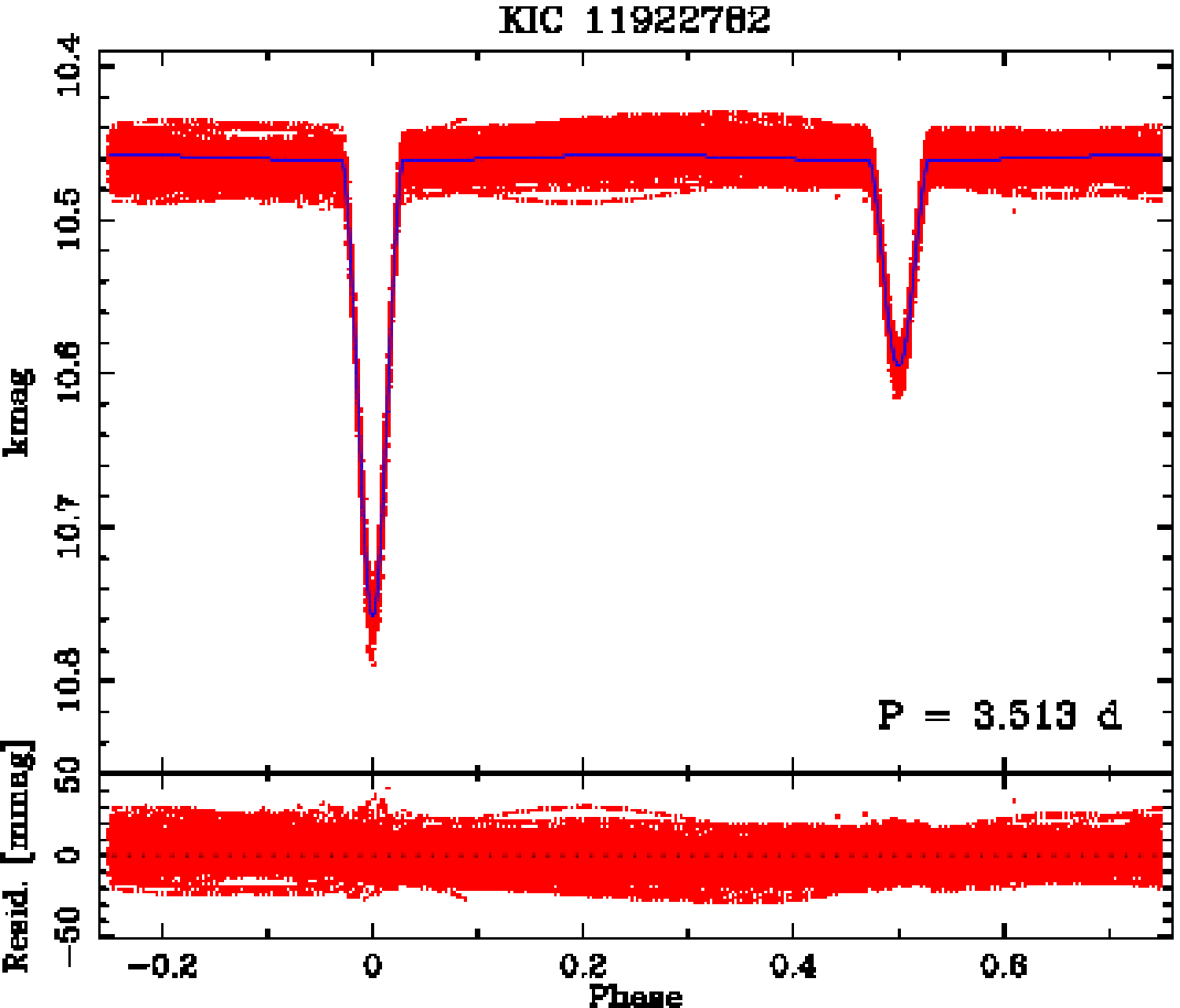}
\caption{Same as Fig. \ref{fig_mod_0343}, but for KIC~11922782.
}\label{fig_mod_1192}
\end{figure*}

This system also has no new HIDES observations, and we kept the results of the
analysis of the complete long-cadence LC (shown in Fig.~\ref{fig_mod_1192}) from Paper~II. 
Short-cadence data are available only for pieces of Q02 and Q03. Similarly to KIC~8552540, the LC
is affected by rapidly evolving spots, so two incomplete \kep quarters of data would not be 
sufficient for proper fitting. Moreover, the depths and widths of the minima in the 
long-cadence curve has not been affected by longer exposure time.

The star is another case of a low flux ratio binary. 
For the record, the precision in masses (from Paper~II) is 0.9 + 0.7
per cent for the primary and secondary, respectively, and 3.8 + 7.4 per cent analogously in radii.
The uncertainty of secondary's fractional radius $r_2$ is mainly affected by strong activity 
and rapid evolution (in a time scale of days) of prominent, cool spots. Notably, secondary is the 
lowest-mass star (0.85~M$_\odot$) in our entire \kep sample of SB2s. The primary, on the other hand is 
an evolved and older analogue of the Sun, with a similar mass but significantly larger radius. 
The whole system is therefore interesting for several reasons.

Spectral analysis with \isp was possible only on the TD spectrum of the primary (SNR=149).
KIC~11922782 was found to be metal-poor with respect to the Sun, which was expected for
an older system. Also $T_{\rm eff,1}=6000(110)$~K, which is higher than solar, is explainable
considering the metal depletion. It is however in disagreement with $5620^{+64}_{-70}$~K 
from GDR2. Rotationally broadened lines point strongly towards synchronous
rotation, which, according to the theory, should be achieved after 20~Myr. The circular orbit is 
also not surprising, as the time scale of circularisation is only about 220~Myr.
Lack of independent estimates of $T_{\rm eff,2}$ prevents us from using \jktabs for distance 
calculations.

\subsection{Age and evolutionary status}\label{sec_iso}

\begin{figure*}
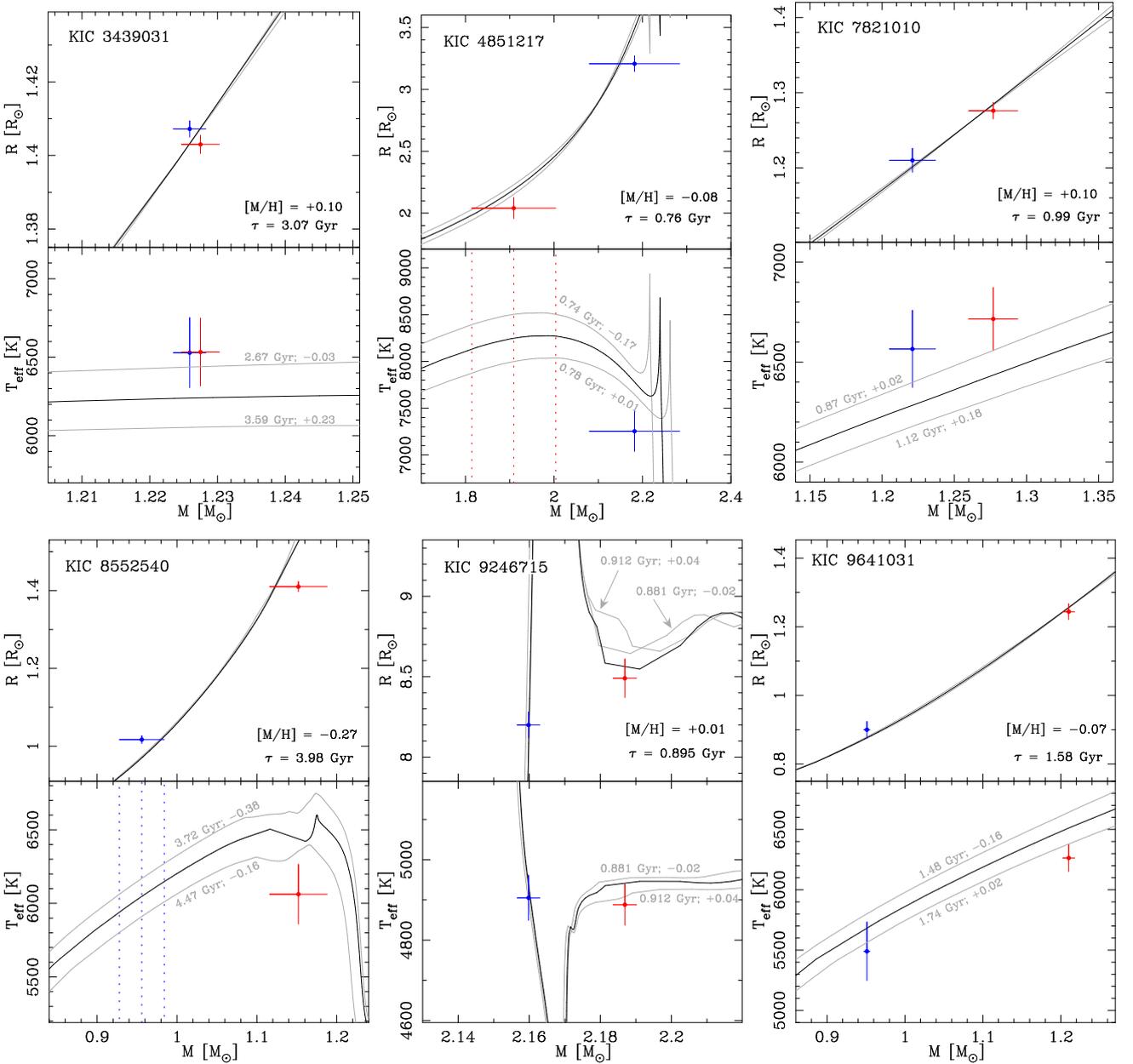

\includegraphics[width=0.32\textwidth]{age_K0343.eps}
\includegraphics[width=0.32\textwidth]{age_K0485.eps}
\includegraphics[width=0.32\textwidth]{age_K0782.eps}
\\ \vspace{0.2 cm}
\includegraphics[width=0.32\textwidth]{age_K0855.eps}
\includegraphics[width=0.32\textwidth]{age_K0924.eps}
\includegraphics[width=0.32\textwidth]{age_K0964.eps}
\caption{Comparison of our results with MESA isochrones on 
$M/R$ (upper) and $M/T_{\rm eff}$ (lower) planes. Primaries are shown with 
red, and secondaries with blue symbols, both with 1$\sigma$ errorbars.
Black lines are isochrones for metallicites adopted from \isp analysis
that best match both components on both planes simultaneously, with [M/H] and 
best-fitting age $\tau$ also labelled in black. When a $T_{\rm eff}$ has 
not been measured, dotted lines of a corresponding colour on the $M/T_{\rm eff}$ 
plane mark the mass and $\pm$1$\sigma$ error of the component, in order
to evaluate its $T^i_{\rm eff}$. Grey lines 
represent isochrones for metallicity varied by $\pm$1$\sigma$, that best
match all available data on both planes. On $M/R$ they are
often undistinguishable from the black line, because of the age-metallicity 
degeneration. Their corresponding $\tau$ and [M/H] are given on 
$M/T_{\rm eff}$ panels. This Figure shows the results for KICs: 3439031, 
4851217, 7821010, 8552540, 9246715, and 9641031.
}\label{fig_iso1}
\end{figure*}

\begin{figure*}
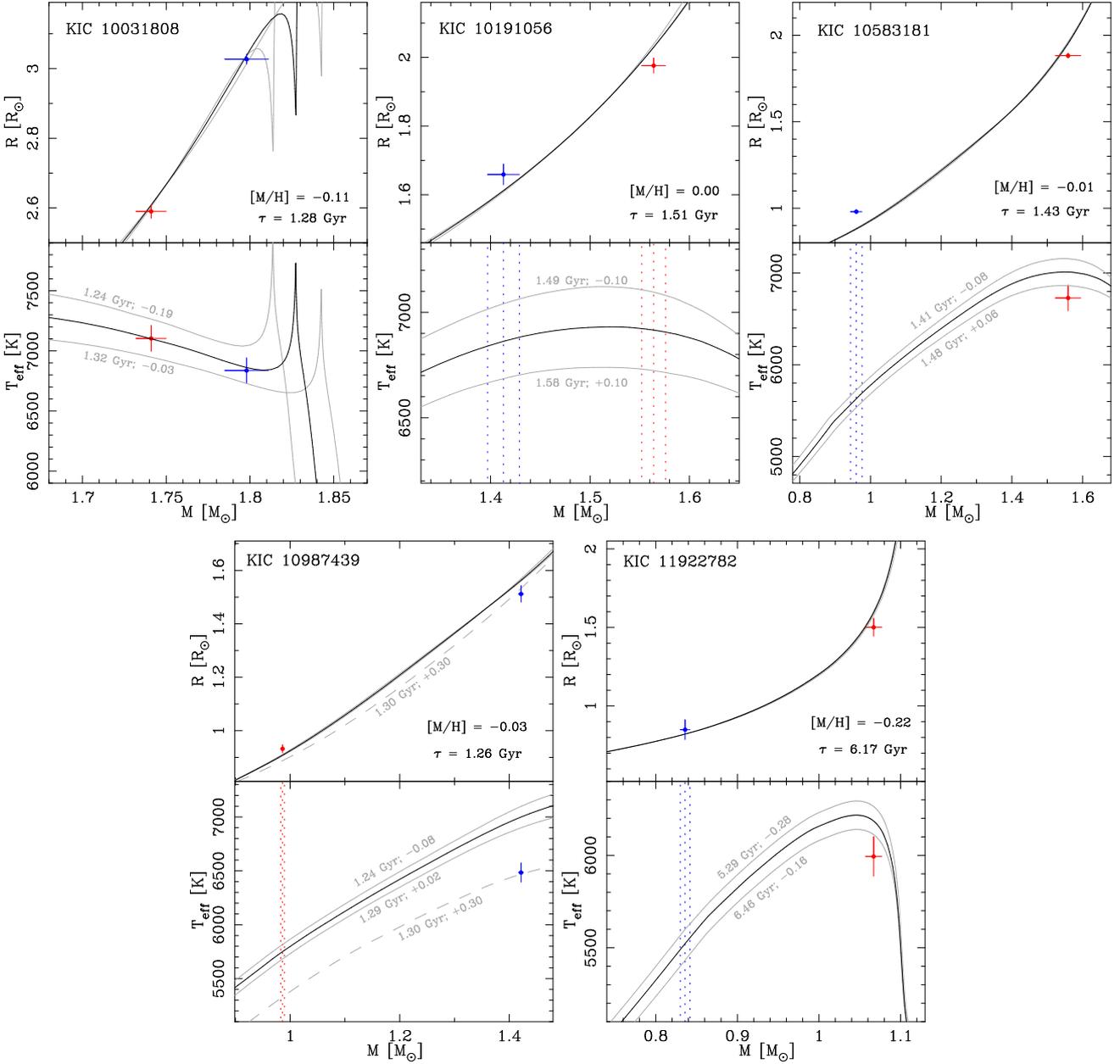

\includegraphics[width=0.32\textwidth]{age_K1003.eps}
\includegraphics[width=0.32\textwidth]{age_K1019.eps}
\includegraphics[width=0.32\textwidth]{age_K1058.eps}
\\ \vspace{0.2 cm}
\includegraphics[width=0.32\textwidth]{age_K1098.eps}
\includegraphics[width=0.32\textwidth]{age_K1192.eps}
\caption{Same as Figure~\ref{fig_iso1} but for KICs: 10031808, 
10191056, 10583181, 10987439, and 11922782. For KIC~10191056
the [M/H] and its uncertainty have been assumed. For KIC~10987439
we additionally plot an $\tau=1.3$~Gyr, [M/H$]=+0.30$~dex isochrone
(dashed line). See text for details.
}\label{fig_iso2}
\end{figure*}

In Figures \ref{fig_iso1} and \ref{fig_iso2} we show the comparison of 
the results of our analyses with theoretical MESA isochrones. For each
system we determine the age, on which the observed or calculated properties
of both components are best represented. Since the stellar mass $M$ is the most 
robust resulting parameter, that also strongly determines the evolution of a star,
we compare our data on mass-radius ($M/R$) and mass-effective temperature 
($M/T_{\rm eff}$) planes. The former is prone to age-metallicity degeneration,
meaning the same $M/R$ combination can be reproduced by various pairs of
[M/H] and $\tau$. On the other hand, the $M/T_{\rm eff}$ plane is relatively
insensitive to changes in $\tau$, when a star is on the MS, but very sensitive 
to [M/H]. Therefore, a combination of $M/R$ and $M/T_{\rm eff}$ planes can be 
used to determine the age and evolutionary status securely, as long as [M/H] and
at least one $T_{\rm eff}$ are estimated.

For systems with only one effective temperature determined with {\sc ispec}, 
we use the best-fitting isochrone, and the dynamical mass estimate for the
faint companion, to check the $T_{\rm eff}$ predicted for the given mass
and metallicity at the given age. We use this isochrone-based $T^i_{\rm eff}$
together with the other $T_{\rm eff}$ as input in \jktabs to evaluate
the distance $d_{\rm J}$. This distance is then compared with a value
from GDR2. In case of KIC~10191056, both temperatures are estimated such way, 
and solar metallicity is assumed.

\subsubsection{KIC 3439031}
This is a system with two nearly identical stars residing on the MS, but 
somewhat evolved with respect to the Zero-Age Main Sequence (ZAMS).
The age of 3.07~Gyr is, as expected, below the upper limit for orbit 
circularisation -- 4.27~Gyr. The agreement of the model with temperatures
is quite good, but mainly due to relatively large errors in $T_{\rm eff}$ 
and [M/H]. Notably, individual values of [M/H] and [$\alpha$/Fe] 
from \isp were almost identical: [M/H] = 0.09(13) and 0.10(13)~dex for the 
primary and secondary, respectively, and [$\alpha$/Fe] = 0.02(14) and 0.03(14)~dex,
analogously. At the same time, the temperatures we adopted led
to $d_{\rm J}$ that agrees very well with GDR2, therefore we suspect that it
is the metallicity scale that is systematically shifted off, rather than the 
temperatures. Even so, the age will still be $>$2.5~Gyr, and the conclusions
about the evolutionary status remain unchanged.

\subsubsection{KIC 4851217}
Only one estimate of $T_{\rm eff}$ from spectra is available for this pair.
We estimated the age of this system to be 760~Myr, and conclude that the more massive 
secondary has nearly finished its MS evolution. From the $M/R$ diagram only,
this system would be younger, but then the temperature of the secondary
would be significantly larger. This component is now at the stage, when $T_{\rm eff}$
changes rapidly, in comparison to earlier MS stages, thus its age determination is
highly sensitive to the temperature. The best-fitting isochrone, for [M/H] found
in {\sc ispec}, matches our parameters at $<$3$\sigma$ level.
Slightly better match is found for higher metallicities, therefore we suspect 
that this system is more metal abundant than what we obtained. It is worth
to remember that the secondary is a fast rotator, which might have affected our 
analysis, but in any case its evolutionary stage is established securely.

The isochrone-predicted effective temperature $T^i_{\rm eff}$ of the fainter 
but hotter primary is 8250(300)~K (uncertainty includes error in [M/H],
see Figure~\ref{fig_iso1}). When used in \jktabs together with $T_{\rm eff,2}$ from 
{\sc ispec}, the predicted distance $d_{\rm J}$ is 1090(70)~pc, assuming $E(B-V)=0.03$~mag
to reach consistency between all bands. The agreement with 1195(53)~pc from 
GDR2 is therefore quite good. This shows that, despite we do not have complete 
information about KIC~4851217, our results are reliable. 

\subsubsection{KIC 7821010}
This pair consists of two MS components, therefore the strongest age constraints
(for a particular value of [M/H]) come from the $M/R$ diagram, while the
$M/T_{\rm eff}$ plane helps to assess the metal content. Assuming the \isp 
value of [M/H] = +0.10~dex, we get the best overall fit for $\tau = 0.99$~Gyr,
with the model matching our results on the mass-radius plane very well,
and within 2$\sigma$ on the $M/T_{\rm eff}$ diagram.
A better match ($\sim$1$\sigma$) is obtained for lower [M/H] = +0.02~dex 
(-1$\sigma$ from the \isp value), for which the best overall fit is found
for $\tau = 0.87$~Gyr. We can therefore securely conclude that KIC~7821010
is younger than the Sun, so the small metal enhancement is not surprising.
Notably, the secondary is similar in mass to both components of KIC~3439031, 
but is significantly smaller (and likely hotter). Also, both systems have similar 
metallicites. All this is consistent with KIC~7821010
being younger, than KIC~3439031 ($\tau\simeq3.1$Gyr).

\subsubsection{KIC 8552540 (V2277 Cyg)}
The primary of KIC~8552540 comes to the end of its MS evolution, which is 
supported by its oversized radius and lower temperature. The best-fitting
isochrone, assuming the \isp value of [M/H$] = -0.27$~dex, is found for 
$\tau \simeq 4.0$~Gyr. Varying the metallicity by its 1$\sigma$ uncertainty,
changes $\tau$ by $0.3-0.5$~Gyr. The ``older'' value, for higher metal
content of $-$0.16~dex, seems to better match the $T_{\rm eff,1}$, but still the 
agreement between our temperature determination from \isp and the prediction
for [M/H$] = -0.27$~dex is formally within 2$\sigma$. As in the case of
KIC~4851217, the component analysed with \isp rotates rapidly ($\sim$67~k\ms),
which might have affected the results of spectral analysis.

The formally best-fitting 3.98~Gyr, $-$0.27~dex isochrone predicts the temperature 
of the fainter secondary to be $T^i_{\rm eff,2}=6050(230)$~K, which is 
almost exactly the same as $T_{\rm eff,1}$. This is clearly in contradiction 
to the fact that the \kep LC shows two very much uneven eclipses (Fig.~\ref{fig_mod_0855}), 
which (for $e=0$) means different surface brightnesses, hence temperatures. 
Even though a formal agreement with isochrones exists, our temperature and/or 
metallicity scales are likely affected, presumably by the rotation, as already
mentioned. Notably a solar-composition model can reproduce our $T_{\rm eff,1}$, and
the $M/R$ diagram quite well for $\tau = 5$~Gyr.

For the distance estimates with \jktabs we used several pairs of isochrone-predicted
temperatures, taken from isochrones ranging from [M/H] = 0.0~dex (5~Gyr) to
$-$0.38~dex (3.72 Gyr). All the resulting distances are significantly above the
value of 231(1)~pc from GDR2, i.e.: 
269~pc for [M/H$] = -$0.38~dex, $T^i_{\rm eff, 1}=6630$~K, $T^i_{\rm eff, 2}=6180$~K, $E(B-V)$=0.09~mag;
267~pc for $-$0.27~dex, 6440~K, 6050~K, 0.06~mag;
266~pc for $-$0.16~dex, 6360~K, 5910~K, 0.03~mag;
and 259~pc for 0.0~dex, 6060~K, 5700~K, and 0.0~mag. 
In all cases we assumed 200~K errors in temperatures, obtained $\sim$9~pc 
uncertainty in distance, and the $E(B-V)$ value was found by forcing 
individual distances from $V,I,J,H,K$ bands to give the lowest
spread\footnote{In this case we used $I$ instead of $B$, because it was 
available in {\it Simbad}, and the $B$ band magnitude was giving distances
very much deviated from the other bands.}.

We could not obtain an \isp fit with the assumed [M/H] higher than our result from 
Table~\ref{tab_par_sb2}. Nevertheless, we conclude that KIC~8552540 is probably 
more metal abundant, and its components hotter than what we have found. Additional 
problems with more reliable age and [M/H] assessment come from the fact that the 
precision in masses is relatively poor, which prevents us from using other 
information (i.e. flux ratio, absolute magnitudes from GDR2 distance) to 
discriminate between various [M/H]$-\tau-T_{\rm eff}$ scenarios. Nevertheless, 
our results are good enough to conclude that both components are still at the MS 
(with the primary approaching its end), and the system is a few Gyr old.

\subsubsection{KIC~9246715}
The exact evolutionary stage of KIC~9246715 was somewhat uncertain. \citet{raw16}
established from asteroseismology that the oscillating star is at the core-He burning
phase (secondary red clump). In their comparison with MESA models, they assumed
higher than typical mixing-length parameter $\alpha=2.5$, in order to explain
the primary's radius smaller than normally expected from a horizontal-branch star.
Since in their solution both components are very similar to each other, they concluded
that both stars are currently on the red clump, although they also considered an 
option where both components are still on the red giant branch (RG; before core-He burning), 
but would have to be of a slightly different age.

Our solution shows that the two masses differ significantly ($>$6$\sigma$), and the 
primary's radius is also slightly larger than obtained by \citeauthor{raw16} (Tab.~\ref{tab_k924}). 
Large enough, in fact, that the secondary can be a horizontal-branch star, without assuming 
an abnormal mixing-length parameter, or weaker convective overshooting \citep[which was also 
discussed by][]{raw16}. We found a very good match on both $M/R$ and $M/T_{\rm eff}$ planes 
for an age of nearly 0.9~Gyr (Fig.~\ref{fig_iso1}). In such a situation the primary is just 
after He ignition in the core ($M>2.18$~M$_\odot$), while the secondary is still on 
the RG phase, growing rapidly. Since this phase lasts very shortly, the 
secondary constrains the age to a precision of only few Myr. An age-metallicity
degeneration is still present, but our [M/H] estimates allow us to reduce the 
overall uncertainty in $\tau$ (for this particular set of models) down to $<$20~Myr.

\subsubsection{KIC 9641031 (FL Lyr)}
The notable differences between our recent solution and the one from \citet{pop86},
are smaller $R_2$, lower [M/H], and larger $T_{\rm eff}$s (Tab.~\ref{tab_fllyr}).
All this results in a much better match to the isochrones than previously. 
As can be seen in Figure~\ref{fig_iso1}, both components are nicely
represented by a 1.58~Gyr, [M/H$]=-0.07$~dex model on the $M/R$ plane, as
well as on the $M/T_{\rm eff}$ plane when [M/H] uncertainty is taken into account.
The agreement on the $M/R$ diagram would not be possible with \citeauthor{pop86}'s
parameters, and their very high metal content (+0.32~dex) was assumed to match the 
effective temperatures, which themselves were derived from now-outdated calibrations, 
and were indirectly dependent on the flux and radius ratio in their solution. 
We claim this work's results to be more credible.

Notably, FL~Lyr is listed in the DEBCat\footnote{\tt http://www.astro.keele.ac.uk/$\sim$jkt/debcat/}, 
which is a collection of the best-studied detached eclipsing binaries \citep{sou15}. 
Even very recently it has been used for creating state-of-the-art calibrations
of fundamental stellar parameters \citep[e.g.][]{eke15,gra17}.
Its case shows, however, that many systems studied several decades ago may need 
a revision, and their parameters need to be updated with better precision and accuracy,
if they are to be used for the purposes of modern astrophysics. 

\subsubsection{KIC 10031808}
Comparison of our results with the MESA isochrones shows an excellent match on both
$M/R$ and $M/T_{\rm eff}$ diagrams for the age of 1.28~Gyr. The overall uncertainty
in $\tau$ is well below 100~Myr (taking into account errors in all parameters), which
makes the estimate of the age of this system especially important for asteroseismology.
KIC~10031808 is one of a very few examples of a well-studied eclipsing binary with a 
$\gamma$~Doradus pulsator, which in this case we suspect is the hotter, less massive 
primary. The cooler secondary could also have been a pulsator in the past, 
but currently may be too evolved for stable pulsation modes. Considering that the two
masses do not differ very much, it is therefore interesting to see a system in which 
both components are close to the end of their MS lifetimes, but (probably) only one is still 
showing pulsations. With a complete set of high-precision and high-accuracy stellar
parameters, and a good age estimate, KIC~10031808 should be a subject of a dedicated, 
detailed asteroseismic study, which we strongly encourage.

\subsubsection{KIC 10191056 A}
In this case, where we have no independent estimates of effective temperatures, 
we decided to repeat the approach from Paper~II, and fit a solar-composition
([M/H]=0.0) isochrone to the $M/R$ only, but using the updated parameters and 
MESA models. We also assumed 0.1~dex as the 1$\sigma$ uncertainty in [M/H] 
to estimate the variations in age and the temperatures. 

The best fit, within $\sim$1$\sigma$ agreement, was found for 
$\tau=1.51$~Gyr. The primary's radius suggests a significant deviation from the 
ZAMS. The predicted values of $T^i_{\rm eff}$ are nearly identical -- 6880(200)
and 6850(180)~K for the primary and secondary, respectively -- which is
consistent with the LC showing nearly equal-depth eclipses. To estimate the distance, 
we followed the same procedure as in Paper~II, where we only used apparent $V$ and
$I$ band magnitudes, corrected for the third light: 11.19(17) and 10.51(5)~mag, 
respectively. These values come from fitting the $V$ and $I$-band light
curves, obtained from ASAS-K (see Paper~II).
The GDR2 distance of 613(8)~pc is reached when $E(B-V)$ is assumed to be 
0.081~mag. In such case the two individual distances are 650(70) and 608(26)
for $V$ and $I$ band, respectively, and their weighted average is 613(31)~pc.
The isochrone-predicted flux ratio in the \kep band is 0.61(17), which agrees 
within errors with 0.68(4) from the LC fit. 

We can also estimate the isochrone-predicted mass of the tertiary component
KIC~10191056~B. For the $l_2/l_1$ value predicted by isochrones, the 
$l_3/l_{\rm tot}$ contribution of 0.160(9) is reproduced by a star of 
$M_{\rm B}=1.23(3)$~M$_\odot$. The GDR2 lists GP magnitudes of both A and B.
The assumed isochrone predicts the total absolute GP magnitude of A 1.73(12)~mag,
and the magnitude difference from GDR2 is 1.923(6)~mag, thus the absolute
GP magnitude of B would be 3.65(12)~mag. This value is reproduced by a star 
of $M_{\rm B}=1.230(25)$~M$_\odot$, which is the same as the previous
value. This consistency suggests that the component B is an F-type star, 
gravitationally bound to the eclipsing pair~A.

\subsubsection{KIC 10583181}
In this system  both primary and secondary are much larger than expected at ZAMS.
While this can be easily explained by evolution for the primary, the reason why
the secondary is oversized must be different. Most likely it is related to 
activity and fast rotation. This is not surprising, as similar or even
higher level of inflation is observed in other 
short-period systems with nearly-solar-mass components, such as for the primary 
of HP~Aur \citep[$P=1.42$~d, $M=0.954$~M$_\odot$, $R=1.028$~R$_\odot$;][]{lac14},
the secondary of ZZ~UMa \citep[$P=2.30$~d, $M=0.972$~M$_\odot$, $R=1.16$~R$_\odot$;][]{lac99}, 
or the secondary of KIC~8552540 ($P=1.06$~d, $M=0.956$~M$_\odot$, $R=1.02$~R$_\odot$; 
Tab.~\ref{tab_par_sb2}). On the other hand, the secondary of KIC~9641031
($P=2.18$~d, $M=0.951$~M$_\odot$, $R=0.900$~R$_\odot$; Tab.~\ref{tab_par_sb2})
does not show such behaviour. The exact reason of the radius discrepancy of KIC~10583181~B
is unclear to us.

We could not find a satisfactory fit to both components simultaneously, 
therefore we decided to search for the age of the system basing on the primary only.
An isochrone for $\tau=1.43$~Gyr nicely reproduces its position on the $M/R$ plane, 
and is in a decent agreement with our $T_{\rm eff}$ estimates. As expected, it
predicts the primary to be near the end of the MS stage. The match is better when
[M/H] error is taken into account, which resembles the situation with KICs
4851217 and 8552540. In all these cases the possible factor affecting the \isp 
analysis is probably the rotational velocity. Nevertheless, the agreement is
still acceptable, and the evolutionary stage is determined securely.

The adopted best-fitting isochrone predicts the secondary to have 
$T^i_{\rm eff}=5630(160)$~K. Unfortunately, because this system is a triple with 
significant flux contribution from the third body, and we have no information
on $l_3$ in bands other than {\it Kepler}'s, we can not use \jktabs to estimate 
the distance. Instead, we calculate the apparent magnitude of the primary, and
compare it to the absolute one, predicted by MESA isochrones. As stated above, 
we suspect $l_3$ was varying slightly throughout the \kep observations, therefore
for the following analysis we take its conservative uncertainty, and assume its 
contribution to be 0.12(2). 

With the primary's contribution to the total flux of 0.796(20), its apparent magnitude 
should be 11.26(3)~mag\footnote{We do not take into account the uncertainty
of the satellite's photometric zero point, but we understand it may introduce
additional source of error.}. The absolute brightness, estimated from the isochrone,
is 2.28(20)~mag, with uncertainty in [M/H] and mass taken into account.
Analogously, for the secondary we obtain the apparent brightness of 13.70(4)~mag,
and absolute of 4.95(18)~mag. The weighted average of the distance module
is therefore 8.85(19)~mag, which translates into distance (without extinction)
of 589(52)~pc. To level it with the GDR2 value of 445(5)~pc, one needs to assume 
$E(B-V)\simeq0.20$~mag. This would require the EW of the sodium D~1 line to 
be around, 0.4~\AA\, but in the spectra we measured it to be only 0.192(16)~\AA\,
\citep[$E(B-V)<0.1$~mag; ][]{mun97}.
We would also like to note that the isochrone-predicted flux ratio $l_2/l_1$
of 0.086(30) agrees with the value of 0.105(3) from the LC fit.

We do not attempt to estimate the mass of the tertiary from isochrones, as 
this object may be a binary itself, and its true contribution to the total
flux in the \kep band is actually uncertain.

A better insight into this system would be possible with additional information
about the secondary and tertiary. The former can come from analysis of a 
higher-SNR disentangled spectrum, which requires additional observations with a 
telescope larger than OAO-188cm, and/or advanced post-processing. The latter
can be achieved with multi-colour photometry, even from the ground. A $\sim$2~mmag
precision observations are certainly possible, and the flat secondary eclipse 
helps to constrain the fluxes in the LC analysis.

\subsubsection{KIC 10987439}
With masses and radii measured to a very high precision, this system poses a 
challenge to isochrone fitting. A good fit (within 2$\sigma$) is achieved on the
$M/R$ plane to both components with an isochrone of $\tau=1.26$~Gyr. The more 
massive secondary, which dominates the total flux of the system, 
does not seem to be very evolved, and the primary not significantly inflated
(similarly to KIC~9641031~B). Unfortunately, the \isp value of $T_{\rm eff,2}$ 
is significantly ($\sim$500~K) lower than 7000(110)~K that the isochrone predicts. 
This can not be explained by the rotation ($v\sin(i)=7.16$~k\ms), nor the SNR of the 
disentangled spectrum ($\sim$135). All the \isp runs, for which the starting 
$T_{\rm eff}$ was set to 7000~K or much higher, converged to values around 
6500~K. Also when the initial [M/H] was set to higher values ($\sim$0.2-0.3~dex)
the result was the same. In an additional check, we used four line depth ratio
(LDR) vs. $T_{\rm eff}$ calibrations from \citet{kov04}, that utilise lines in
the available wavelength regions and have the smallest $rms$ (below 50~K).
In this way we obtained the average $T^L_{\rm eff}=6380(100)$~K, supporting
the \isp result. This discrepancy between models and \isp is puzzling,
however, we believe that the key is the metallicity scale, which might have
been affected because of the shortest wavelength range, on which the analysis
was performed. The blue end of the TD spectrum is at 5470~\AA, while in case of
other stars it is at 5030~\AA\,(except KIC~7821010). 

We get a very good match to 
the $T_{\rm eff,2}$ value on the MS for [M/H]=+0.30~dex isochrones. 
The formally-best fit was found for the age of $\sim$1.3 Gyr. We plot this model 
in Fig.~\ref{fig_iso2} as well, for comparison. The flux ratios $l_2/l_1$ 
predicted by the isochrones are 7.00(25) and 6.37(77) for [M/H]=+0.30 and 
-0.03~dex, respectively. While the latter is much closer to the value obtained
from LC fit -- 6.5(1.1) -- both are in formal agreement within error bars.

The effective temperatures predicted by the [M/H$]=-0.03$~dex, 
$\tau=1.26$~Gyr isochrone are $T_{\rm eff,1} = 5750(80)$~K and 
$T_{\rm eff,2} = 7000(110)$~K. The distance obtained with \jktabs for these
values is 339(11)~pc, under the assumption of $E(B-V)=0.1$~mag, made to
reach the consistency between various bands. When the \isp value of 
$T_{\rm eff,2}$ is used, the distance is 333(13)~pc with $E(B-V)=0.0$~mag.
Finally, $T^i_{\rm eff,1}$ predicted by the [M/H$]=+0.30$~dex, 
$\tau=1.30$~Gyr isochrone is 5330(80)~K. Together with $T_{\rm eff,2}$
from {\sc ispec}, it leads to the distance of 326(11)~pc at $E(B-V)=0.1$~mag.
We measured the EW of the interstellar sodium D~1 line and found it to be
0.16(1)~\AA. Which, according to calibrations by \citet{mun97}, corresponds
to $E(B-V)\sim$0.05~mag. This favours the cases with the $T_{\rm eff,2}$
from {\sc ispec}. All values are significantly lower than 374(4)~pc from GDR2.
In any case, both components of KIC~10987439 are currently at the MS, and 
the age of the systems seems to be $\sim$1.3~Gyr. With the gDor-type
pulsations detected from the dominant (secondary) star, which has its
parameters well constrained, KIC~10987439 is worth further detailed studies.
It offers an interesting insight into the mechanisms of pulsations and their
stability conditions, especially considering that the pulsator seems to reside 
at the low-mass edge of the gDor instability strip.

\subsubsection{KIC 11922782}
This system shows one of the lowest metallicities, and is also the oldest.
The best-fitting [M/H$] = -0.22$~dex isochrone was found for $\tau = 6.17$~Gyr.
The primary, which is similar in mass to the Sun, but much larger, is 
currently at the end of its MS life. Notably, the 6.17~Gyr isochrone matches
both components on the $M/R$ plane very well, which is not always observed
in stars of mass similar to the secondary. The agreement with measured
$T_{\rm eff,1}$ is also quite good, within 2$\sigma$, and even better when 
[M/H] error is taken into account. In terms of masses and metallicity, 
KIC~11922782 is similar to V636~Cen \citep[$1.052+0.854$~M$_\odot$, 
$-$0.20~dex][]{cla09}. Also the radii of 
secondaries are comparable: 0.830~R$_\odot$ in V636~Cen vs. 0.849~R$_\odot$
in this work. The age found for V636~Cen is $\sim$1.4~Gyr, assuming different
mixing length scales for the two components. The much larger primary's
radius in KIC~11922782 (1.501~R$_\odot$ vs. 1.018~R$_\odot$ in V636~Cen)
confirms the older age. Additionally, we did not have to fine-tune the
mixing length parameters to reach agreement with models.

The best-fitting isochrone predicts $T^i_{\rm eff,2} = 5520(120)$~K.
When used in \jktabs for distance determination, it results in 
$d_{\rm J} = 227(12)$~pc, under the assumption of $E(B-V)=0.09$~mag. 
This is in a good agreement with 236(2)~pc from GDR2.

\subsection{KIC~10191056~Bb -- a candidate M dwarf}\label{sec_k1019b}

\begin{table}
\centering
\caption{New HIDES RV measurements of KIC~10191056~B}\label{tab_rv_k1019b}
\begin{tabular}{lccrc}
\hline \hline
BJD & $v$ & $\epsilon$ & $t_{\rm exp}$ & SNR \\
-2450000 & (k\ms ) & (k\ms ) & (s) & \\
\hline
6852.954108$^a$	& -22.329	& 0.387	&  480	& 16	\\
7815.288473	& -24.192	& 0.257	& 1500	& 31	\\
7817.272376	& -24.292	& 0.175	& 1500	& 48	\\
7845.230234	& -24.143	& 0.127	& 1500	& 29	\\
7893.090781	& -24.027	& 0.178	& 1500	& 30	\\
7894.155920	& -24.399	& 0.242	& 1800	& 25	\\
7949.014405	& -24.458	& 0.151	& 1630	& 45	\\
7955.090230	& -24.011	& 0.096	& 1511	& 50	\\
8035.000297	& -24.568	& 0.121 & 1588	& 46	\\
\hline
\end{tabular}
\\$^a$ From TRES spectrum from 2014.07.14.
\end{table}

\begin{table}
\centering
\scriptsize
\caption{Parameters of orbital solutions to the RVs of KIC~10191056~B for different 
(fixed) eccentricities, and mass of the host $M_{\rm B}=1.23(3)$~M$_\odot$.
Uncertainties are formal fit errors.}\label{tab_v3_bd}
\begin{tabular}{cccccc}
\hline \hline
$e$	& $P$ & $K$ & $a\sin(i)$ & $f(m)$ & $m\sin(i)$ \\
(fix)& (d) & (k\ms) & (AU) & (M$_J$) & (M$_J$) \\
\hline
0.0	&  2180$\pm$380	& 1.087$\pm$0.140	& 0.22$\pm$0.01	& 0.30$\pm$0.05	&  80$\pm$11 \\
0.1	&  2350$\pm$1320& 1.116$\pm$0.328	& 0.25$\pm$0.06	& 0.35$\pm$0.20	&  83$\pm$30 \\
0.2	&  2800$\pm$1990& 1.171$\pm$0.436 	& 0.28$\pm$0.13	& 0.46$\pm$0.33	&  91$\pm$40 \\
0.3	&  3680$\pm$1450& 1.187$\pm$0.514	& 0.35$\pm$0.06	& 0.58$\pm$0.24	&  99$\pm$45 \\
0.4	&  4920$\pm$1330& 1.183$\pm$0.535	& 0.41$\pm$0.05	& 0.70$\pm$0.21	& 104$\pm$48 \\
0.5	&  6750$\pm$1720& 1.177$\pm$0.539	& 0.47$\pm$0.05	& 0.78$\pm$0.24	& 109$\pm$51 \\
0.6	&  9760$\pm$2640& 1.174$\pm$0.539	& 0.54$\pm$0.06	& 0.88$\pm$0.29	& 113$\pm$53 \\
0.7	& 15500$\pm$4530& 1.172$\pm$0.538	& 0.61$\pm$0.07	& 0.99$\pm$0.35	& 118$\pm$57 \\
0.8	& 29280$\pm$9130& 1.172$\pm$0.538	& 0.68$\pm$0.09	& 1.10$\pm$0.44	& 122$\pm$58 \\
\hline
\end{tabular}
\\The $rms$ of a fit varies from 196~\ms\, for $e=0.0$ to 189~\ms\, for $e=0.8$.
\end{table}

\begin{figure}
\includegraphics[width=0.9\columnwidth]{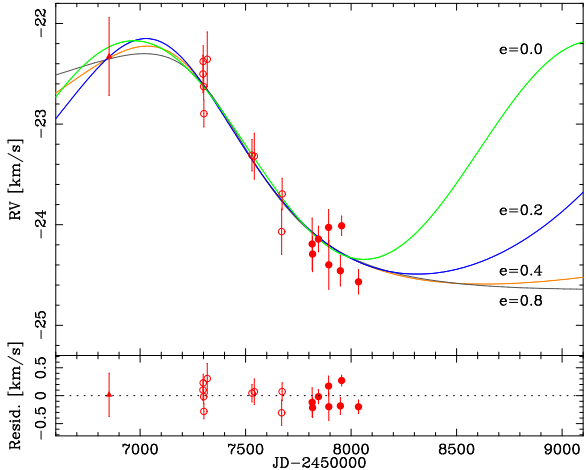}
\caption{RV measurements of KIC~10191056~B, and examples of orbital solutions for 
selected eccentricities (labelled). Old HIDES, new HIDES, and TRES data
are plotted with open circles, filled circles, and a triangle, respectively.
The lower panel shows residuals for $e=0.8$, but other
solutions give residuals that are practically indistinguishable.
}\label{fig_v3_bd}
\end{figure}

In Paper~II we presented two hypotheses about the origin of RV variations of 
KIC~10191056~B. First, that we observe motion on the orbit around common 
centre of mass with the eclipsing pair A. Second, that we see a short-period (10-20~d)
modulation induced by a massive planet or low-mass brown dwarf (BD). The new 
HIDES observations, supported with the TRES spectrum from July 2014, make the former
very unlikely, and rule out the latter. New observations do not match any of the
previously proposed solutions. The motion of B is, however, very clear. 

In Figure~\ref{fig_v3_bd} we show all the RV measurements for KIC~10191056~B. 
The new ones are presented in Table~\ref{tab_rv_k1019b}.
New HIDES data (after JD=2457800) cluster at lower values, 
while the TRES point is at a similar level as the earliest ones from HIDES
(no zero-point shift was assumed, see Sect.~\ref{sec_1019A}). We clearly
see a gradual drop in velocity, with some curvature. Since we do not see a
significant long-term upwards trend in $\gamma$ of the pair A, this observed
RV drop must be due to another body in the system (Bb), that is orbiting B with a
period much longer than the time span of our observations.

With only a fraction of the orbit covered, we can not fit for $P$ and $e$
simultaneously. Instead, we run a number of fits with fixed eccentricity
that varied from 0.0 to 0.8. Their results are shown in Table~\ref{tab_v3_bd}.
Solutions with $e>0.8$ lead to very long orbital period and showed problems 
with convergence, therefore we decided not to present them. Nevertheless, 
such high eccentricities are not impossible. The circular ($e=0$) solution 
sets the lower limit on the period of Bb at 2180(380)~d. All solutions have
similar quality, with $rms\simeq190$~\ms. 

Independently of the assumed $e$, all solutions lead to RV amplitude of 
$\sim$1.1~k\ms, which strongly constrains the minimum mass of the companion. 
We obtained $m\sin(i)$ between 80 and 122 Jupiter mass (M$_J$), with large
uncertainties coming mainly from the unknown $P$, and assuming a theoretically-predicted
mass of the host 1.23(3)~M$_\odot$. This means that
the companion is most likely a late-M type star, or, rather unlikely, a massive 
BD. More observations are needed to set further limits on orbital parameters
and $m\sin(i)$. Even a few observations taken throughout 2019 will allow to distinguish 
between low- and high-eccentricity solutions. The spectra need to be of a sufficiently
high resolution, and taken around quadratures of the pair A, to have the narrow lines 
of B distinguishable and well separated from broad lines of the pair A. Unfortunately,
the available observations from LAMOST \citep[e.g.; ][]{fra16,zon18}, or those used by
\citet{mat17}, do not have sufficient spectral resolution. A better estimate of 
$M_{\rm B}$ would also be welcome.

\section{Conclusions}\label{sec_conc}
We presented updated results for 11 (3 new) DEBs from the original \kep satellite
observing field, based on new observation, refined light curve fits, 
and/or additional, complementary methods of analysis. We improve our knowledge on 
the systems, mainly by adding new information, crucial for assessing the age and 
evolutionary status of the studied targets. In this work we studied a variety of 
interesting systems, including a double-giant pair with one of the components at the core He
burning phase, a probable quadruple, various pairs with components at vastly 
different phases of MS evolution, three DEBs with pulsating (dSct and 
gDor) components, a pair of twins, and a pair with a low-mass (0.85~M$_\odot$)
secondary. A significant number of systems have their physical absolute 
parameters derived with $<$2~per cent uncertainties. For two targets -- 
KIC 9246715 and KIC~9641031 (FL~Lyr) -- our updated results are in better 
consistency with evolutionary models than in the previous studies. 
There are still open questions and things to improve in some cases.
First, the light curves could be cleaned from influence of spots or pulsations,
and a more careful de-trending could be performed. The effective temperatures 
of fainter components could be derived, but this requires more spectroscopic observations and
higher SNR spectra. We also encourage the community to study the pulsations
in presented systems.
 
\section*{Acknowledgements}
We would like to thank: 
Sergi Blanco-Cuaresma from the Harvard-Smithsonian Center 
for Astrophysics, the creator of {\sc ispec}, for making available the version of 
his code before it was made public; 
the organisers and presenters of the 2018 iSpec 
Spectroscopic Summer School in Wroc{\l}aw, especially Ewa Niemczura (Astronomical Institute, University 
of Wroc{\l}aw), Barry Smalley (Keele University Astrophysics Group), and S.B.-C. for 
fruitful discussions, tips, suggestions, and comments regarding the spectral analysis;
and Miroslav Fedurco from the Pavol Josef \v{S}af\'arik University in
Ko\v{s}ice, Slovakia, for the discussion of results.

This research has made use of the SIMBAD database, operated at CDS, Strasbourg, France.
This work has made use of data from the European Space Agency (ESA)
mission {\it Gaia} (\url{http://www.cosmos.esa.int/gaia}), processed by
the {\it Gaia} Data Processing and Analysis Consortium (DPAC,
\url{http://www.cosmos.esa.int/web/gaia/dpac/consortium}). Funding
for the DPAC has been provided by national institutions, in particular
the institutions participating in the {\it Gaia} Multilateral Agreement.

KGH, MR and AP acknowledge support provided by the Polish National Science Center through grants no. 2016/21/B/ST9/01613, 2015/16/S/ST9/00461, and 2016/21/B/ST9/01126 respectively. This work was partially supported by JSPS KAKENHI Grant Number 16H01106.








\appendix

\section{New observations of KIC~4758368}\label{sec_app_k4758}

\begin{table}
\centering
\caption{New HIDES RV measurements of KIC~4758368}\label{tab_rv_k475}
\begin{tabular}{lccrc}
\hline \hline
BJD & $v$ & $\epsilon$ & $t_{\rm exp}$ & SNR \\
$-$2450000 & (k\ms ) & (k\ms ) & (s) & \\
\hline
7892.148808	&	$-$31.176	&	0.040	& 1200	& 37	\\
7948.177541	&	$-$31.394	&	0.051	&  900	& 20	\\	
\hline
\end{tabular}
\end{table}

\begin{table}
\centering
\caption{New and previous orbital solutions and parameters of KIC~4758368.}\label{tab_k475}
\begin{tabular}{lcc}
\hline \hline
Parameter		& Paper~I & This work \\
\hline
{\it Outer orbit} &&\\
$P_{\rm AB}$~(d)		& 2553(80)	& 5030(1160) \\
$T_0$ (BJD-2450000)		& 5581(9)	& 5570(7) \\
$K_{\rm A}$ (k\ms)		& 8.81(41)	&  8.72(37) \\
$K_{\rm B}$ (k\ms)		&12.9(1.4)	& 16.8(2.8) \\
$e_{\rm AB}$			& 0.672(37)	& 0.80(6) \\
$\omega_{\rm A}$ ($^\circ$)& 142(2)	& 134(6) \\
$\gamma$ (k\ms)			&$-$24.0(1.2)& $-$29.4(1.9) \\
$M_{\rm A} \sin^3(i)$ (M$_\odot$)	& 0.65(18)	& 1.2(7)\\
$M_{\rm B} \sin^3(i)$ (M$_\odot$)	& 0.45(9) 	& 0.63(32)\\
$a_{\rm AB}\sin(i)$ (AU)& 3.77(32)	& 9.3(2.1) \\
$N_{\rm A}$		& 106	& 106	\\
$N_{\rm B}$		& 19	& 21	\\
$rms_{\rm A}$ (k\ms)	& 6.9	& 6.9	\\
$rms_{\rm B}$ (\ms)	& 57    & 63 \\
\hline
\jkt {\it solution} &&\\
$P_{\rm A}$ (d)$^a$	& \multicolumn{2}{c}{3.74993552(43)} \\
$r_{\rm Aa} + r_{\rm Ab}$	& \multicolumn{2}{c}{0.4252(64)} \\
$r_{\rm Ab}/r_{\rm Aa}$  	& \multicolumn{2}{c}{1.22(21)}\\
$l_{\rm B}/l_{\rm tot}$		& \multicolumn{2}{c}{0.725}\\
\hline
{\it Absolute values} &&\\
$d_{\rm GDR2}$ (pc)		&  \multicolumn{2}{c}{1910(80)}\\
$M_{\rm B}$ (M$_\odot$)	&  \multicolumn{2}{c}{1.43$^b$}\\
$i_{\rm AB}$ ($^\circ$)	&    43(5)	&  50(16) \\
$M_{\rm A}$ (M$_\odot$)	&   2.1(6)	& 2.7(1.6)\\
$a_{\rm AB}$ (AU)		&  5.54(72)	& 12(4) \\
$\hat{a}_{\rm AB}$ (mas)&  2.9(4)	& 6.4(2.0) \\
$a_{\rm A}$ (R$_\odot$)	&  13(1)	& 14(3)\\
$R_{\rm Aa}$ (R$_\odot$)&  2.48(35)	& 2.7(7) \\
$R_{\rm Ab}$ (R$_\odot$)&  3.04(27)	& 3.3(8) \\
\hline
\end{tabular}
\\$^a$Orbital period of the eclipsing inner pair.
\\$^b$From the KIC values of $\log(g)$ and $R$ given without errors.
\end{table}

\begin{figure}
\includegraphics[width=0.9\columnwidth]{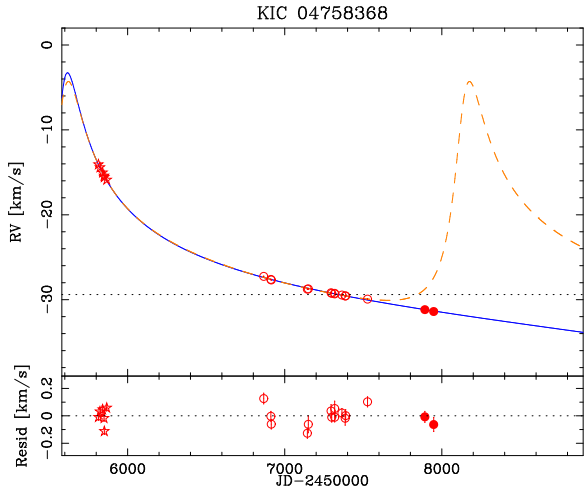}
\caption{Directly measured RVs of KIC~4758368~B (red symbols), 
with two orbital solutions: from Paper~I (orange dashed line) and
this work (blue solid). The APOGEE, old HIDES, and new HIDES data
are marked with open stars, open circles, and filled circles, 
respectively. The post-ETV velocities of the inner pair A were 
also used in the fit, but we omit them in the Figure for clarity.
}\label{fig_k475}
\end{figure}

Apart from the sample of double- and triple-lined spectroscopic systems, we
re-observed KIC~4758368 (KOI~6448), which only shows one set of prominent lines, 
and was previously described in Paper~I. This target is a hierarchical triple, 
with a short-period ($P\simeq3.75$~d) eclipsing binary, orbited by a third star
on an eccentric, long-period orbit. The third star, which is a red giant, 
is the dominant source, and we see its lines in spectra. 
In Paper~I we used our 13 HIDES measurements and
6 APOGEE data points for the third star, and a set of RVs of the centre of
mass of the inner DEB, obtained from its eclipse timing variations (post-ETV RVs). 
In that work, our data coverage was too short to securely establish the
outer period, thus we observed this target in May and July 2017, recording
two more spectra. The previous solution predicted a small but measurable rise
of the RVs, which would help to constrain the period. Instead, we continued to
observe a downward trend, which clearly means the outer period is much longer 
than what we have obtained previously. The new measurements are listed in 
Table~\ref{tab_rv_k475}.

As in Paper~I, we treated the system as a double-lined binary, with post-ETV
RV measurements for one component (designated A), and directly measured RVs 
for the other (designated B). 
We followed the same fitting procedure with \vfit as previously. 
Table~\ref{tab_k475} and Figure~\ref{fig_k475} show comparison of old
and new solutions. The most obvious difference is the much longer orbital period, 
and other parameters dependent on it ($a_{\rm AB}\sin(i)$, $M_{\rm A,B} \sin^3(i)$).
Large relative uncertainty in $P$ also leads to large uncertainties in other 
parameters.

As in Paper~I, we assumed the mass of B to be 1.43~M$_\odot$ (taken from the 
\kep Input Catalog) and \jkt results of LC fit (details in Paper~I), 
and estimated other values, such as: inclination of the outer orbit $i_{\rm AB}$,
its absolute ($a_{\rm AB}$) and angular ($\hat{a}_{\rm AB}$) major 
semi-axes, total mass of the inner pair $M_{\rm A}$, and absolute radii
of the inner pair components $R_{\rm Aa,Ab}$. We conclude, that the inner
pair is likely composed of two sub-giant stars, slightly more massive than
the Sun, although the uncertainties are very large. Interestingly, 
the orbital motion of B around common centre of mass is strong enough
to be detectable astrometrically with AO facilities.

There are other estimates of  $M_{\rm B}$ in the literature, reaching up to 
2.09~M$_\odot$. For such scenario, the inner pair's total mass would be
$\sim$4.0~M$_\odot$, and its components radii would be 3.0-3.7~R$_\odot$,
making them comparable in flux or even brighter than the apparently 
dominant star B. We thus find it unlikely. Other $M_{\rm B}$ estimates
are below solar mass (0.7-0.85~M$_\odot$), but then the system would have
to be significantly older than 10~Gyr for B to be a giant.

\section{New RV measurements of SB2s}

In Table~\ref{tab_RV} we present individual RV measurements used
in this work, that were not previously published. TRES and APOGEE data
are properly distinguished. For both components
of a given SB2 we show the measured RV values $v_{1,2}$, their errors
$\epsilon_{1,2}$ (both in k\ms), as well as exposure times in seconds 
and SNR at $\lambda\sim5500$~\AA\ for optical spectra, 
and $\lambda\sim12000$~\AA\, for the IR.

\begin{table*}
\centering
\caption{All RVs of SB2 pairs used in this work.
Complete Table is available in the on-line version of the manuscript.}\label{tab_RV}
\begin{tabular}{lcccclrlc}
\hline \hline
BJD & $v_1$ & $\epsilon_1$ & $v_2$ & $\epsilon_2$ & KIC & $t_{\rm exp}$ & SNR & Note$^a$ \\
-2450000 & (k\ms ) & (k\ms ) & (k\ms ) & (k\ms )  & & (s) & \\
\hline
7673.020510 & 100.172 & 0.131 & $-$46.325 & 0.219 &  3439031 & 1500 & 24 &\\
7812.327593 & $-$50.460 & 0.137 & 104.593 & 0.098 &  3439031 & 1500 & 29 &\\
7846.336225 &  60.889 & 0.199 &  $-$6.204 & 0.210 &  3439031 & 1500 & 25 &\\
...\\
5823.726555 & $-$122.76 & 6.00 &  54.11 & 3.96 &  4851217 & 3386 & 119 & A \\
5849.578427 &   48.68 & 5.13 & $-$81.25 & 5.63 &  4851217 & 2002 & 142 & A \\
5851.648819 &  $-$79.27 & 4.62 &  13.59 & 6.14 &  4851217 & 2002 & 134 & A \\

... \\
6867.030540 &  $-$46.473 & 0.092 &   13.313 & 0.145 &  7821010 & 1800 & 48 &\\
6869.142604 &  $-$38.627 & 0.361 &    5.871 & 0.532 &  7821010 & 1500 & 23 &\\
6914.079695 &  $-$51.604 & 0.102 &   18.896 & 0.132 &  7821010 & 1500 & 63 &\\
... \\
\hline
\end{tabular}
\\$^a$ Spectra taken with facilities other than OAO-188cm/HIDES are marked:
  ``A'' for APOGEE, and ``T'' for TRES. 
\end{table*}


\bsp	
\label{lastpage}
\end{document}